\documentstyle[psfig,aps,preprint]{revtex}
\tightenlines
\pssilent

\newcommand{\hs}{\hat{s}^2}
\newcommand{\hc}{\hat{c}^2}
\newcommand{\ori}{\hspace{.1in}}
\newcommand{\fp}{f^{\prime}}
\newcommand{\ti}{\tilde}
\makeatletter
\def\slash{\@ifnextchar[{\fmsl@sh}{\fmsl@sh[0mu]}}
\def\fmsl@sh[#1]#2{%
  \mathchoice
    {\@fmsl@sh\displaystyle{#1}{#2}}%
    {\@fmsl@sh\textstyle{#1}{#2}}%
    {\@fmsl@sh\scriptstyle{#1}{#2}}%
    {\@fmsl@sh\scriptscriptstyle{#1}{#2}}}
\def\@fmsl@sh#1#2#3{\m@th\ooalign{$\hfil#1\mkern#2/\hfil$\crcr$#1#3$}}
\makeatother
\def\lsim{\mathrel{\rlap {\raise.5ex\hbox{$ < $}}
{\lower.5ex\hbox{$\sim$}}}}
\def\gsim{\mathrel{\rlap {\raise.5ex\hbox{$ > $}}
{\lower.5ex\hbox{$\sim$}}}}

\def\Re{{\rm Re}\, }
\def\np#1#2#3{{\it {Nucl. Phys.}} {\bf{B#1,}} #2 (#3)}
\def\pl#1#2#3{{\it {Phys. Lett.}} {\bf{B#1,}} #2 (#3)}
\def\prl#1#2#3{{\it {Phys. Rev. Lett. }}{\bf{#1,}} #2 (#3)}
\def\pr#1#2#3{{\it {Phys. Rev.}} {\bf{D#1,}} #2 (#3)}
\def\zp#1#2#3{{\it {Z. Phys.}} {\bf{C#1,}} #2 (#3)}

\newcommand{\mpl}[2]{{\em Mod. Phys. Lett.}     {\bf A#1}, #2 }
\begin{document}
\begin{titlepage}
\begin{flushright}

hep-ph/9801425{\hskip.5cm}\\
\end{flushright}
\begin{centering}
\vspace{.3in}
{\bf  THE EFFECTIVE WEAK MIXING ANGLE IN THE MSSM
}\\
\vspace{2 cm}
{A. DEDES$^{1}$,
A. B. LAHANAS$^2$ and K. TAMVAKIS$^1$}\\
\vskip 1cm
{\it $^1$Division of Theoretical Physics,}\\
{\it University of Ioannina, GR-45110, Greece}\\ \vspace{0.5cm}
{\it $^2$Physics Department, Nuclear and Particle Physics Section
University of Athens, Athens 157 71, Greece}\\ \vspace{0.5cm}

\vspace{1.5cm}
\begin{abstract}
The predictions of the MSSM are discussed in the light of recent LEP and SLD 
precision data. The full supersymmetric one loop corrections to the effective 
weak mixing angle, experimentally determined in LEP and SLD experiments, are 
considered. It is demonstrated, both analytically and numerically, that,
 potentially 
dangerous, large logarithmic sparticle corrections are cancelled. The relative 
difference factor $\Delta k$ between the mixing angle defined as a ratio of
 couplings and the experimentally obtained angle is discussed. It is found 
 that $\Delta k$ is dominated by the oblique corrections, while the non-oblique 
 overall supersymmetric EW and SQCD corrections are negligible. The comparison 
 of the MSSM with radiative electroweak symmetry breaking to the LEP+SLD
 precision 
 data indicates that rather large values of the soft breaking parameter $M_{1/2}$ 
 in the region greater than 500 GeV are preferred.
\end{abstract}

\end{centering}
\vspace{.1in}

\vspace{1cm}
\begin{flushleft}

January 1998\\
\end{flushleft}
\hrule width 6.7cm \vskip.1mm{\small \small}
E-mails\,:\,adedes@cc.uoi.gr,
 alahanas@atlas.uoa.gr, tamvakis@cc.uoi.gr

\end{titlepage}

\section { INTRODUCTION}

The electroweak mixing angle $\sin^2\theta_{W}$ , defined as a ratio of 
gauge couplings, provides a convenient means to test unification in
unified extensions of the Standard Model (SM) \cite{Lang}. This
quantity is not directly measured in experiments. Instead, LEP and SLD
studies employ an effective coupling  $\sin^2\theta^{lepton}_{eff}$ 
determined from on resonance asymmetries whose value is known with excellent
accuracy  \cite{LEP,altarelli,sirlin,gambino}.
 This effective mixing angle has been studied in detail in the
context of the SM at the one loop level in various renormalization schemes
with the dominant two loop heavy top contributions and three loop QCD
effects taken into account \cite{sirlin,gambino,all1,all2,djoua}.
Due to large cancellations between fermion and
boson contributions occurring at the one loop level,
in the $\overline{MS}$ scheme, these are the dominant
contributions to the difference 
 $ \sin^2\theta^{lepton}_{eff} -  \sin^2\theta_{W} \approx {\cal{O}}
  {(10^{-4})} $ which is less than the error quoted by the experimental
groups. Therefore, although conceptually different the two angles are
very close numerically.
The mixing angle  is sensitive on the values of the Higgs mass
$M_{H}$ and top mass $m_{t}$  through the quantities $\Delta {r_{W}}$
and $\Delta {\rho}$ and carries an uncertainty of about $.1 \%$
from its dependence on the electromagnetic coupling ${\alpha} (M_{Z}) $.
>From the predictions of $ \sin^2\theta^{lepton}_{eff} $
and $\Delta {\rho}$ one can draw useful theoretical conclusions concerning
the Higgs and W - boson masses having as inputs the Z - boson mass, the
value of the fine structure constant and the Fermi coupling constant which are
experimentally known to a high degree of accuracy.

In the framework of supersymmetric extensions of the SM \cite{NHK}
 the situation changes
since $\sin^2\theta_{W}$ as well as $ \sin^2\theta^{lepton}_{eff} $
receive contributions from the superparticles in addition to ordinary
particles. Coupling unification at the GUT scale in conjunction with
experimental data for the strong coupling constant at $M_{Z}$ and radiative
breaking of the Electroweak Symmetry impose stringent constraints on the
extracted value for $\sin^2\theta_{W}$. However $\sin^2\theta_{W}$
is plagued by large logarithms  $\log(M_Z/M_S)$, where $M_S$ is the
effective supersymmetry breaking scale\footnote{See 
for instance P. Chankowski, Z. Plucienic and S. Pokorski in 
ref.\cite{polon}.}. Unlike $\sin^2\theta_{W}$ the
experimentally determined $ \sin^2\theta^{lepton}_{eff} $  is not plagued
by such potentially dangerous large logarithms due to decoupling.
Therefore, the difference of the two angles is not numerically small
any more and $\sin^2\theta_{W}$ cannot be directly used for comparison
with experimental data. Thus, in supersymmetric theories the precise relation
between the two angles is highly demanded.
The non-decoupled supersymmetric corrections to $\sin^2\theta^{lepton}_{eff}$
are expected to be small of order $(M_Z/M_S)^2$.
However small these contributions may be,
they are of particular importance, since the
experimental accuracy is very high, and these corrections can be larger than
the SM corrections occurring beyond the one loop order.
Moreover the effect of the one loop supersymmetric corrections
may not be necessarily suppressed in some sectors, such as the
neutralino and chargino sectors, which are
characterized by a relatively small effective supersymmetry breaking scale for
particular inputs of the soft SUSY breaking parameters.
Motivated by this we undertake a complete one loop
study of the supersymmetric corrections to the effective
mixing angle in the context of the MSSM which is the simplest supersymmetric
extension of the Standard Theory.

Although there are several studies \cite{polon} in literature concerning
the value of the weak mixing angle $\sin^2\theta_{W}$ in the MSSM and other
unified supersymmetric extensions of the SM,
only a few have tackled the
problem of calculating the complete supersymmetric corrections
${\cal {O}} {(M_Z/M_S)^2}$ to
the experimentally measured angle $ \sin^2\theta^{lepton}_{eff} $.
 In ref.\cite{Bagger}  the effective mixing angle is calculated in 
particular cases and the 
decoupling of large logarithms is numerically shown.
In that calculation all the one-loop corrections, including the non-universal
supersymmetric vertex and external fermion corrections , for the leptonic
effective mixing angle were considered. The non-universal corrections were
found to be small.
In other studies \cite{Hollik1,finnell,kolda},
the serious constraints 
imposed by unification and radiative electroweak symmetry breaking
\cite{Tamvakis} have not been considered. Instead the MSSM parameters are
considered as free parameters chosen in the optimal way to improve
the observed deficiencies of the SM in describing the data.
 
In the present article we show explicitly how the cancellation 
of potentially dangerous logs takes place and perform a systematic
numerical study by scanning the entire parameter space having as our main
outputs the effective weak mixing angle, the values of the on Z - resonance
asymmetries measured in experiments, as well as the value of the strong
coupling constant at $M_Z$. In each case we also give the theoretical
prediction for the W - boson mass through its relation to the parameter rho
and the weak mixing angle.

It is perhaps worth noting that non-universal
corrections, claimed to be small, are dominated by large logs.
These logs cancel at the end, as expected. Nevertheless, their presence
dictates that non-decoupled terms of order $(M_Z/M_S)^2$ may be of the
same order of magnitude as the corresponding terms stemming from
the universal corrections and cannot be a priori omitted.
Knowing from other studies that universal corrections tend to
decrease the value of the effective mixing angle by almost six
standard deviations from the experimental central value it is
important to see what is the effect of the non-universal
contributions. We take into account
all constraints from unification and radiative
EW symmetry breaking. These constraints, along with the
experimental bounds for the strong coupling constant and
$\sin^2\theta^{lept}_{eff}$, may restrict further the allowed parameter space. 

\section{FORMULATION OF THE PROBLEM}

The value of the weak mixing angle, defined as the ratio of
the gauge couplings, is  
%%%%%%%%%%%%%%
\begin{equation}
\hs(Q) = \frac{\hat{g}^{\prime 2}(Q)}{\hat{g}^2(Q)+\hat{g}^{\prime 2}(Q)}  \;,
\label{sinratio}
\end{equation}
%%%%%%%%%%%%%%%%%
where $\hat{g}$ and $\hat{g}^\prime$ are the $SU(2)$ and $U(1)_Y$
gauge couplings. Throughout this paper the hat refers to 
renormalized quantities in the modified $\overline{DR}$ scheme
\cite{Siegel,Martin}.
These couplings are running in the sense that they
depend on the scale Q. Particularly for the electroweak processes,
Q is chosen to be $M_Z$. There are many sources for the determination
of the $\hs$. From muon decay, for instance, and knowing that 
$M_Z=91.1867 \pm 0.0020 GeV$, $\alpha_{EM}=1/137.036$ and 
$G_F=1.16639(1)\times10^{-5}\, GeV^{-2}$, we get in the ($\overline{DR}$)
scheme
%%%%%%%%%%%%%%
\begin{equation}
\hs \hc=\frac{\pi\, \alpha_{EM}\, 
}{\sqrt{2}\,M_Z^2\,G_F\,
(1-\Delta \hat{\alpha})\,\hat{\rho}\,(1-\Delta \hat{r}_W)}\; ,
\label{sindr}
\end{equation}
%%%%%%%%%%%%%%%%%
where 
%%%%%%%%%%%%%%%%%
\begin{eqnarray}
\hat{\rho}^{-1}&=&1-\Delta \hat{\rho}=
1-\frac{\Pi_{ZZ}(M_Z^2)}{M_Z^2}+\frac{\Pi_{WW}(M_W^2)}
{M_W^2}\label{ro}\;,\\[3mm]
\Delta \hat {r}_W&=&\frac{\Pi_{WW}(0)-\Pi_{WW}(M_W^2)}{M_W^2}+
\hat{\delta}_{VB}\label{drw}\;,\\[3mm]
\hat{\alpha}&=&\frac{\alpha_{EM}}{1-\Delta{\hat{\alpha}}}
\label{da}\;.
\end{eqnarray}
%%%%%%%%%%%%%%%%%%%%
$\Pi$'s are the transverse gauge bosons self energies 
evaluated in the $\overline{DR}$ scheme.
Explicit forms for these self energies can be obtained from
ref.\cite{Bagger}.
The weak mixing angle obtained from (\ref{sindr})
although it plays a crucial role in the analysis of
grand unification, it is {\it not}
an experimental quantity. Actually, it is obtained after
fitting experimental observations with
$\alpha_{EM}$ and $G_F$ as accurately known
parameters (for more details {\it see} ref.\cite{Peskin}).
The radiative corrections on $\hs$ involve two subtleties:
{\it i}) the renormalization scheme dependence\footnote{
We are working on the modified $\overline{DR}$ scheme of ref.\cite{Martin}
which preserves supersymmetry up to two-loops.}
and {\it ii}) the 
dependence on the mass of the top quark, Higgs masses and superparticle
masses which depend on the 
supersymmetric breaking parameters $M_{1/2}$, $M_0$, 
and $A_0$.

%\vspace*{0.4in}
%%%%%%%%%%%%%%%%%%%%%%%%%%%
%\centerline{\input{sinhat.ps}}
%\begin{center}
%\footnotesize{{\bf Figure 1:}
% The values of the running weak mixing angle $\hat{s}^2$ at
%$M_Z$ in the $\overline{DR}$ scheme,
% as it is obtained from the ratio of gauge couplings for various
%input universal soft gaugino masses $M_{1/2}$
% which are taken at $M_{GUT}$.
%$\hat{s}^2(M_Z)$ starts strongly increasing
% for $M_{1/2}>>M_Z$ or $M_{1/2}<<M_Z$
%due to the non-decoupling of SUSY-particles
% from its expression (\ref{sindr}).}
%\end{center}
%%%%%%%%%%%%%%%%%%%%%%%%%%%

 As we can see from  Fig.1, 
$\hs$ takes large values
when we increase  the masses of the soft breaking parameters.
In other words, the soft breaking parameters do not decouple
from $\hs$. This
is due to the fact that the net effect of the contributions
(\ref{ro}),(\ref{drw}),(\ref{da}) to (\ref{sindr}),
 contains large logarithms
of the form $\log(\frac{M_{SUSY}}{M_Z})$. 
On the other hand, the LEP collaborations \cite{LEP} employ an
{\it effective weak mixing angle} $\sin^2\theta^{f}_{eff}
\equiv s^2_f$, first introduced
by the authors of ref.\cite{sirlin}, which is not plagued by large logarithms
due to decoupling.
It is a common belief among
GUT theorists that these two angles $\hs$ and $s_f^2$, although
different conceptually,  are very close numerically \cite{gambino}. 
Nevertheless, this is not true in the MSSM since there are 
large logarithmic dependencies of the weak mixing angle $\hs$.

The tree level Lagrangian associated with the $Zf\overline f$
can be written in the form
%%%%%%%%%%%%%%
\begin{equation}
{\cal L}_{tree}^{Zf\overline f}=\frac{\hat{e}}{2\hat{c}\hat{s}}
Z_\mu \overline{f} \gamma^\mu \left [ \left ( T_3^f -2 \hs Q^f 
\right ) - \gamma_5 T_3^f \right ] f \;,
\label{ltree}
\end{equation}
%%%%%%%%%%%%
where $Q^f$ is the electric charge and $T_3^f$ is the third component
of isospin of the fermions $f$. Electroweak corrections in (\ref{ltree})
yield the effective Lagrangian
%%%%%%%%%%%%%%
\begin{equation}
{\cal L}_{eff}^{Zf\overline f}=(\,\sqrt{2}\, G_F 
\, M_Z\,) \, \rho_f^{1/2} Z_\mu \overline{f} \gamma^\mu 
\left [ \left ( T_3^f -2 \hs \hat{k}_f Q^f \right )
 - \gamma_5 T_3^f \right ] f \;,
\label{leff}
\end{equation}
%%%%%%%%%%%%%%%%
which is relevant to study Neutral Current processes on the Z - resonance.
Then, the effective weak mixing angle is simply defined from (\ref{leff}) as
%%%%%%%%%%%%%%
\begin{equation}
s_f^2 \equiv \hs \hat{k}_f = \hs ( 1 + \Delta \hat{k}_f )\;.
\label{sineff}
\end{equation}
%%%%%%%%%%%%%%%%%
The angle $s_f^2$ can be compared directly
with experiment while $\hs$ can be predicted from a Grand
Unification analysis. The LEP and SLD average gives the value
$0.23152\pm 0.00023$ \cite{altarelli} for the  $s_l^2 \equiv
sin^2\theta^{lept}_{eff}$. Since 
%%%%%%%%%%%%%%%%%
\begin{eqnarray}
c_f^2 = \hc \left ( 1 - \frac{\hs}{\hc} \Delta k_f \right )\;,
\nonumber 
\end{eqnarray}
%%%%%%%%%%%%%
one obtains by making use of equations (\ref{sindr}) and 
(\ref{sineff})
%%%%%%%%%%%%%%
\begin{equation}
s^2_{f}c^2_{f}=\frac{\pi\, \alpha_{EM}\, (1+\Delta \hat{k}_f)\,
(1-\frac {\hs}{\hc}\,\Delta  \hat{k}_f)}{\sqrt{2}\,M_Z^2\,G_F\,
(1-\Delta \hat{\alpha})\,\hat{\rho}\,(1-\Delta \hat{r}_W)}\;,
\label{sef}
\end{equation}
%%%%%%%%%%%%%%%%%
where 
%%%%%%%%%%%%%%%%%
\begin{eqnarray}
\Delta \hat{k}_f&=&\frac{\hat{c}}{\hat{s}}\,\frac{\Pi_{Z\gamma}(M_Z^2)-
\Pi_{Z\gamma}(0)}{M_Z^2}+ \frac{\hat{\alpha} \hc}{\pi \hs}\log(\frac{M_W^2}
{M_Z^2}) - \frac{\hat{\alpha}}{4\pi \hs} V_f (M_Z^2) + \delta k_f^{SUSY} \;.
\label{dkapaf}
\end{eqnarray}
%%%%%%%%%%%%%%%
The function $V_f(M_Z^2)$ can be obtained from 
ref.\cite{sirlin}\footnote{In the case f=bottom,
 the important top quark corrections
to $Zb\bar{b}$ vertex should be added to $V_f$.}. 
$\delta k_f^{SUSY}$ denotes the non-universal supersymmetric self energies
and vertex corrections to $s_f^2$. 

In order to study MSSM (or SM) corrections to $s_f^2$, we
need calculate first the $Z$ and $W$ gauge boson self
energy corrections which contribute to 
$\hat\rho$ and $\Delta \hat r_W$. Our expressions agree with those of
ref.\cite{Bagger}\footnote{To match our conventions with those of ref.$\cite{Bagger}$ we have to
replace their matrices by the following: $N\rightarrow {\cal{O}}^T$, and
$U\rightarrow U^{*}$.} and \cite{Rociek}.  
We need also calculate the $Z-\gamma$ propagator corrections, 
the wave function renormalization of external fermions as well as the 
$Z \overline{f} f$ vertex corrections which contribute to $\Delta \hat k_f$.
The supersymmetric contributions to last two were found to be negligible,
 for the leptonic case,
in the minimal supergravity model studied in ref.\cite{Bagger}.
Including all these corrections in (\ref{sef}), we expect that
the effective weak mixing angle
$s_f^2$, does not suffer from potentially large logarithms,
$\sim log(\frac{M^2_{SUSY}}{M_Z^2})$.

At this point we should say that when the electroweak symmetry
is broken by radiative corrections, the value of the parameter $\mu$, which
specifies the mixing of the two Higgs multiplets within the superpotential,
turns out to be  of the order of the supersymmetry breaking scale
in most of the parameter space. Under these circumstances it is not only
the large logarithms $log(\frac{M^2_{SUSY}}{M_Z^2})$ which should be
cancelled but also logarithms involving the parameter $\mu$.

\section{Decoupling of $log(M_{SUSY}/M_Z)$ in the effective mixing angle }

In this section we will first show how the potentially dangerous 
$\sim log(\frac{M_{1,2}}{M_Z}), log(\frac{\mu}{M_Z})$
from the contributions of the neutralinos and charginos are cancelled in
the expression for ${\sin^2}\theta^{f}_{eff}$ when the soft SUSY breaking
parameter $M_{1/2}$ is large ($M_{SUSY}^2>>M_Z^2$).

There are three sources of large logarithms which affect the value of the
weak mixing angle $\sin^2\theta^{f}_{eff}$  :\\
i)   Gauge boson self energies which feed large logs to the quantities
$\Delta \hat r_W$, $\hat\rho$ and $\Delta \hat k_f$.  \\
ii)  Vertex, external wave function renormalizations and box corrections
to muon decay which affect $\Delta \hat r_W$ through $\delta_{VB}^{SUSY}$. \\
iii) Non-universal vertex and external fermion corrections to 
$Z \overline{f} f$ coupling which affects $\Delta \hat k_f$. \\
We shall see that the corrections
 (i) are cancelled against large logs stemming from the
electromagnetic coupling ${\hat{\alpha}}(M_Z)$. The rest, (ii) and (iii),
are cancelled against themselves.

In order to prove the cancellation of the large $log(M_{SUSY}/M_Z)$ terms
among the dimensionless quantities $\Delta \hat r_W$, $\hat\rho$,
$\Delta \hat k_f$ and ${\hat{\alpha}}(M_Z)$, through which
$\sin^2\theta^{f}_{eff}$ is defined, it suffices
to ignore the electroweak
symmetry breaking effects {\it e.g} $<H_1^o>=<H_2^o>=0$. In this
case the masses of charginos and neutralinos take
the simple form
%%%%%%%%%%%%%
\begin{eqnarray}
m_{{\chi}_i^o}&=&M_1\,\,\,,\,\,\,M_2\,\,\,,\,\,\,\mu\,\,\,,\,\,\,-\mu \;,\\
m_{{\chi}_i^{+}}&=&M_2\,\,\,,\,\,\,\mu\;.
\end{eqnarray}
%%%%%%%%%%%%%%%%%%%%%%%

\vspace*{1cm}
{\bf{i) \underline{Vector boson self energy corrections }}}

\vspace*{.4cm}
 The contributions  from the chargino/neutralino sector to the
vector bosons self energies are
%%%%%%%%%%%%%
%%%%%%%%%%%%%
\begin{eqnarray}
\Pi_{ZZ}^{\chi_i^o/\chi_i^+}&=&\frac{\hat{g}^2}{16\pi^2}
\Biggl \{\frac{1}{2}
H(\mu,\mu)\left [\frac{1}{\hat{c}^2}+4\left (\hat{c}-\frac{1}{2\hat{c}}
\right )^2\right ]+
\mu^2B_0(\mu,\mu)\left [\frac{1}{\hat{c}^2}+4\left 
(\hat{c}-\frac{1}{2\hat{c}}\right )^2 \right ] 
\nonumber \\[2mm] &+& 
2 \hat{c}^2H(M_2,M_2)+4\hat{c}^2M_2^2B_0(M_2,M_2)\Biggr \}\;,\\[5mm]
\Pi_{WW}^{\chi_i^o/\chi_i^+}&=&\frac{\hat{g}^2}{16\pi^2}\Biggl [H(\mu,\mu)+
2\mu^2B_0(\mu,\mu)+2H(M_2,M_2)+4M_2^2B_0(M_2,M_2)\Biggr ]\;,\\[5mm]
\Pi_{Z\gamma}^{\chi_i^o/\chi_i^+}&=&\frac{ \hat{e}
 \hat{g} \hat{c}_{2\theta_W}}
{16\pi^2\hat{c}}
\Biggl [4\tilde{B}_{22}(\mu,\mu)+p^2B_0(\mu,\mu)\Biggr ]\nonumber\\[2mm] &+&
\frac{2 \hat{e}
 \hat{g} \hat{c}}{16\pi^2}\Biggl [4\tilde{B}_{22}(M_2,M_2)+p^2
B_0(M_2,M_2)\Biggr ]\;,
\end{eqnarray}
%%%%%%%%%%%%%%%%%%%%%%%%
where $\hat{g}=\frac{\hat{e}
}{\hat{s}}$ is the running $\overline{DR}$ $SU(2)$ gauge coupling.

In order to calculate the dependence of $s_f^2$ on $M_{1,2} /\mu $ 
we make use of eqs.\ref{ro},\ref{drw},\ref{da} and
reduce all functions appearing in the expressions for the two point
functions above in terms of the basic integrals $A_0, B_0$ (see Appendix B).
Isolating the logarithmic dependencies on $M_{1,2} /\mu $ we find that,
%%%%%%%%%%%%%
\begin{eqnarray}
\Delta
 \hat{r}_W&=&\frac{\hat{\alpha}}{4\pi}
\frac{2}{3\hs}\Biggl [1+2\log\biggl (\frac{M_2^2}{Q^2}\biggr )+
\log\biggl(\frac{\mu^2}
{Q^2}\biggr ) \Biggr]
\;,\\[3mm]
\Delta
 \hat{\rho}&=&\frac{\hat{\alpha}}{4\pi}
\frac{2}{9\hc}
\Biggl [1+2 \hat{c}_{2\theta_W}+6\hc \log\biggl (\frac{M_2^2}{Q^2}\biggr )
%\nonumber \\[1.5mm] &+&3
+3 \hat{c}_{2\theta_W}\log\biggl (\frac{\mu^2}{Q^2}\biggr )\Biggr ]
\;,\\[3mm]
\Delta
 \hat{k}_f&=&-\frac{\hat{\alpha}}{4\pi}
\frac{2 \cot\theta_W}{9\hs}\Biggl [\hat{s}_{2\theta_W}+
\hat{c}_{2\theta_W}\tan\theta_W\nonumber \\[1.5mm] &+&
3\hat{s}_{2\theta_W}
\log\biggl (\frac{M_2^2}{Q^2}\biggr )+3\hat{c}_{2 \theta_W}\tan\theta_W
\log\biggl (\frac{\mu^2}{Q^2}\biggr )\Biggr ]\;,\\[3mm]
\Delta
 \hat{\alpha}&=&-\frac{\hat{\alpha}}{3\pi}\Biggl [\log\biggl (\frac{M_2^2}
{Q^2}\biggr )+\log\biggl (\frac{\mu^2}{Q^2}\biggr )\Biggr ]\;.
\end{eqnarray}
%%%%%%%%%%%%%%%%%%%%%%%
%%%%%%%%%%%%%%%%%%%%%%%
The angle ${\hat{\theta}}_W$ is the weak mixing angle defined through
 ratios of
couplings in the $\overline{DR}$ scheme and 
$\theta_W$ is the on shell mixing angle defined by
$ \sin^2\theta_W = 1-M_W^2/M^2_Z$. In the equations above
$\hat{c}_{2\theta_W} \equiv \cos(2{\hat{\theta}}_W)$,
$\hat{s}_{2\theta_W} \equiv \sin(2{\hat{\theta}}_W)$
with similar definitions for ${c}_{2\theta_W}, {s}_{2\theta_W}$.

Plugging in all that into (\ref{sef}),
 we find that $s^2_{eff}c^2_{eff}$ is corrected
as 
\begin{equation}
\Delta (s^2_{eff}c^2_{eff})=\frac{\pi\, \alpha_{EM}}
{\sqrt{2}\,M_Z^2\,G_F}\,\,\frac{8}{9}(\frac{\hat{\alpha}}{4\pi})\;,
%\delta_{\hat{s}_o^2\hat{c}_o^2}^{M_2}
%=\frac{4}{9}(\frac{\hat{\alpha}}{4\pi})\,\,\,+``non-oblique''
\end{equation}
%%%%%%%%%%%%%%%%%%
which at one loop order is independent of large logs.
It must be noted that this
result is also independent of the sign of $\mu$.
However this finite correction vanishes when the next to
leading terms in the expansion of $B_0$ are considered.

%\subsection{Vertex+Box Corrections}
\vspace*{1cm}
{\bf{ii) \underline{Vertex and Box Corrections from muon decay }}}

\vspace*{.4cm}
The non-universal contribution to $\Delta\hat{r}_W$, which
contains vertex and box as well as external wave
function renormalization corrections, is divided into two parts
%%%%%%%%%%%%
\begin{equation}
\delta_{VB}=\delta_{VB}^{SM}+\delta_{VB}^{SUSY}\;.
\end{equation}
%%%%%%%%%%%%%%%%
The Standard model part appears in ref. $\cite{sirlin} $.  
The supersymmetric contributions can
be found in refs. $\cite{Bagger}$,$\cite{Pokorski}$. We reproduce the
results of ref. $\cite{Bagger}$ for the wave - function and vertex corrections
here,
%%%%%%%%%%%%
\begin{equation}
\delta_{VB}^{SUSY}=-\frac{\hs\hc}{2\pi \hat{\alpha}}M_Z^2{\cal R}e a_1
+(\delta \upsilon_e+\frac{1}{2}\delta Z_e+ \frac{1}{2}\delta Z_{\nu_e})+
(\delta \upsilon_{\mu}+\frac{1}{2}\delta Z_{\mu}+
\frac{1}{2}\delta Z_{\nu_{\mu}})\;,
\label{dvb}
\end{equation}
%%%%%%%%%%%%%%%%
where the wave-function and vertex corrections are
\begin{equation}
16\pi^2~\delta Z_{\nu_e} \ =\ -\ \sum_{i=1}^2
\left|a_{\tilde\chi_i^+\nu_e\tilde e_L}\right|^2
B_1(0,m_{\tilde\chi^+_i},m_{\tilde e_L}) - \sum_{j=1}^4
\left|a_{\tilde\chi_j^0\nu_e\tilde\nu_e}\right|^2
B_1(0,m_{\tilde\chi^0_j},m_{\tilde\nu_e})~,
\end{equation}
\begin{equation}
16\pi^2~\delta Z_e\ =\ -\ \sum_{i=1}^2
\left|a_{\tilde\chi_i^+e\tilde\nu_e}\right|^2
B_1(0,m_{\tilde\chi^+_i},m_{\tilde\nu_e}) \ -\ \sum_{j=1}^4
\left|a_{\tilde\chi_j^0e\tilde e_L}\right|^2
B_1(0,m_{\tilde\chi^0_j},m_{\tilde e_L})~,
\end{equation}
%%%%%%%%%%%%%%%%%%%
\begin{eqnarray}
16\pi^2~\delta v_e &=& \sum_{i=1}^2\sum_{j=1}^4
a_{\tilde\chi_i^+\nu_e\tilde e_L}a^*_{\tilde\chi_j^0e\tilde e_L}
\ \Biggl\{ - \ {\sqrt{2}\over g}a_{\tilde\chi_j^0\tilde\chi_i^+W}
m_{\tilde\chi_i^+} m_{\tilde\chi_j^0} \,C_0(m_{\tilde
e_L},m_{\tilde\chi_i^+},m_{\tilde\chi_j^0}) \nonumber\\
&&\qquad\qquad+ \ {1\over\sqrt2g}b_{\tilde\chi_j^0\tilde\chi_i^+W}
\ \biggl[\,B_0(0,m_{\tilde\chi_i^+},m_{\tilde\chi_j^0}) + m_{\tilde
e_L}^2\,C_0(m_{\tilde e_L},m_{\tilde\chi_i^+},m_{\tilde\chi_j^0}) -
{1\over2}\,\biggr] \Biggr\}\nonumber\\ &-& \sum_{i=1}^2\sum_{j=1}^4
a_{\tilde\chi_i^+e\tilde\nu_e}a_{\tilde\chi_j^0\nu_e\tilde\nu_e}
\ \Biggl\{ - \ {\sqrt2\over g}b_{\tilde\chi_j^0\tilde\chi_i^+W}
m_{\tilde\chi_i^+} m_{\tilde\chi_j^0} \,C_0(m_{\tilde\nu_e},
m_{\tilde\chi_i^+}, m_{\tilde\chi_j^0}) \nonumber\\ &&\qquad\qquad+
\ {1\over\sqrt2g} a_{\tilde\chi_j^0\tilde\chi_i^+W}
\ \biggl[\,B_0(0,m_{\tilde\chi_i^+},m_{\tilde\chi_j^0}) +
m_{\tilde\nu_e}^2 \, C_0(m_{\tilde\nu_e}, m_{\tilde\chi_i^+},
m_{\tilde\chi_j^0}) - {1\over2} \,\biggr]\Biggr\} \nonumber\\ &+&
{1\over2}\sum_{j=1}^4 a_{\tilde\chi_j^0e\tilde e_L}^*
a_{\tilde\chi_j^0\nu_e\tilde\nu_e} \ \biggl[\,B_0(0,m_{\tilde
e_L},m_{\tilde\nu_e}) +
m_{\tilde\chi_j^0}^2\,C_0(m_{\tilde\chi_j^0},m_{\tilde
e_L},m_{\tilde\nu_e}) + {1\over2} \,\biggr]~,
\end{eqnarray}
%%%%%%%%%%%%%%%%%%%%%%%
and the non-vanishing couplings are given by
{\footnote{ To conform with the notation of ref. \cite{Bagger}
we use the couplings
$a_{\tilde\chi_a^0\tilde\chi_i^+W} \equiv  g {\cal P}^L_{a i}$,
$b_{\tilde\chi_a^0\tilde\chi_i^+W} \equiv  g {\cal P}^R_{a i} $. 
$ {\cal P}^L_{a i}$ and ${\cal P}^R_{a i}$ are given in the
Appendix A ( see Eqs. A.13 ).
Also the  lepton, slepton, chargino (or neutralino) couplings in the
equations (25-28) differ in sign
from those given in A.20.}}
%%%%%%%%%%%%%%%%%%%%%%%%%%%%%%%%%%%%%%%%%%
\begin{eqnarray}
a_{\tilde{\chi}^{+}_1\nu_e\tilde{e}_L} &=&
a_{\tilde{\chi}^{+}_1 e \tilde{\nu}_e}=\frac{\hat{e}}{\hat{s}}\;, \\
a_{\tilde{\chi}^{0}_1\nu_e\tilde{\nu}_e}&=&
a_{\tilde{\chi}^{0}_1 e \tilde{e}_L}=-\frac{\hat{e}}{\sqrt{2}
\hat{c}}\;,\\
a_{\tilde{\chi}^{0}_2\nu_e\tilde{\nu}_e}&=&
- \,
a_{\tilde{\chi}^{0}_2e \tilde{e}_L}=\frac{\hat{e}}{\sqrt{2}\hat{s}}\;.
\end{eqnarray}
%%%%%%%%%%%%%%%%%%%%%%%
In all expressions above the functions $B_{0,1}, C_0$ are considered with 
vanishing momenta squared and their analytic expressions in terms of
the masses involved are given in Appendix B. 
%{\footnote{our $B_1$ differs
%in sign from that used in ref. \cite{Bagger}.}.
We recall that we have ignored
EW symmetry breaking effects so that $m_{{\tilde\chi_1}^0}=M_1$,
$m_{{\tilde\chi_2}^0}=M_2$ and $m_{\tilde{\nu}_e}=m_{\tilde{e}_L}=
M_{\tilde{L}}$. We have compared these results with
those of Ref.$\cite{Pokorski}$ and we have found agreement.
Dangerous large log corrections are contained only in the second and
third part of the eq.(\ref{dvb}). For these terms we obtain, 
%%%%%%%%%%%%%%%%%
\begin{eqnarray}
& &\left ( \delta \upsilon_e +\frac{1}{2}\delta Z_e+
 \frac{1}{2}\delta Z_{\nu_e} \right )
= -\frac{1}{\hs}\, \left (\frac{\hat{\alpha}}{4\pi} \right )
\Biggl \{ 2 M_2^2 C_0(M_{\tilde{L}},M_2,M_2) 
\nonumber \\[2mm] &-&  {M_{\tilde{L}}^2} C_0(M_{\tilde{L}},M_2,M_2)
+\frac{1}{4} M_2^2 C_0(M_2,M_{\tilde{L}},M_{\tilde{L}})
-\frac{1}{4} M_1^2 \frac{\hs}{\hc} C_0(M_1,M_{\tilde{L}},M_{\tilde{L}})
\nonumber \\[2mm] &-& 
 B_0(0,M_2,M_2) + \frac{1}{4}\left (1-\frac{\hs}{\hc}
\right ) B_0(0,M_{\tilde{L}},M_{\tilde{L}}) + \frac{1}{2} +
\frac{1}{8} \left ( 1-\frac{\hs}{\hc} \right )
\nonumber \\[2mm] &+& 
\frac{3}{2} B_1(0,M_2,M_{\tilde{L}}) + \frac{\hs}{2\hc}
B_1(0,M_1,M_{\tilde{L}}) \Biggr \} \;.
\label{cdvb}
\end{eqnarray}
%%%%%%%%%%%%%%%%%%%%%
Using Eqs. (B7-B10) of Appendix B, we find that the expression above
involves no large logarithms. Also as said before 
the first term ($\Re\alpha_1$) in eq. (\ref{dvb})
contains finite parts which go as $\sim \frac{M_Z}{M_{SUSY}} $.
Thus no large logarithmic terms arising from the wave function and
vertex corrections of the muon decay and
the decoupling of large logarithms in $s_f^2$ appear.

\vspace*{1cm}
{\bf{iii) \underline{Non-universal corrections to $\Delta\hat{k}_f$}}} 
 
%\subsubsection{ Formulation }
\vspace*{.4cm} 
The $Zf\overline{f}$ vertex corrections can be written as
%%%%%%%%%%%%%%%%%%%%%%
\begin{eqnarray}
i \ori {\frac{\hat{e}}{2 \hat{s} \hat{c}}} \; {\gamma^{\mu}} \;
( \, F_V^{(f)} \, - \, \gamma_5  F_A^{(f)} )\;,
\label{dkf}
\end{eqnarray}
%%%%%%%%%%%%%%%%%%%%%%
where  $ F_V^{(f)} , F_A^{(f)} $ denote the vector and axial couplings
respectively. Incorporating the tree level couplings 
we can write this vertex in a slightly different form as
{\footnote {We follow the notation of Ref.\cite{finnell} which will be useful
in what will follow.}}
%%%%%%%%%%%%%%%%%%
\begin{equation}
i \ori {\frac{\hat{e}}{\hat{s} \hat{c}}} \; {\gamma^{\mu}} \;
( \, u'_L {\cal P_L} + u'_R {\cal P_R} )\;,
\end{equation}
%%%%%%%%%%%%%%%%%%
where
%%%%%%%%%%%%%%%%%
\begin{eqnarray}
u'_L &=& u_L + \frac{F^{(f)}_L}{16\pi^2}\;,\\[2mm]
u'_R &=& u_R + \frac{F^{(f)}_R}{16\pi^2}\;.
\end{eqnarray}
%%%%%%%%%%%%%%
In the equations above $u_L, u_R$ are the tree level left and right handed
couplings respectively related to the vector $v_f$ and axial $a_f$ tree level
couplings by $v_f=u_L+u_R$, $a_f=u_L-u_R$. $ F_{L,R}^{(f)}$ denote
the corresponding one loop corrections to the aforementioned couplings,  
with the coefficient $1/{16\pi^2}$ factored out for
convenience. These are related to $ F_V^{(f)} , F_A^{(f)} $ of eq. (30) by
\begin{eqnarray}
F^{(f)}_V &\equiv&  \frac{1}{16\pi^2} ({F^{(f)}_L}+{F^{(f)}_R})\;, \\[2mm]
F^{(f)}_A &\equiv&  \frac{1}{16\pi^2} ({F^{(f)}_L}-{F^{(f)}_R})  \;.
\end{eqnarray}
%%%%%%%%%%%%%%
As a result the corrections to $\Delta\hat{k}_f$ are given by
%%%%%%%%%%%%%%
\begin{equation}
\Delta\hat{k}_f = - \frac{1}{16\pi^2}\;
\frac{1}{\hs Q_f (u_L-u_R)}\; (\, u_L F_R^{(f)} - u_R F_L^{(f)} \,)\;,
\end{equation}
%%%%%%%%%%%%%%%
and are equivalent to the well known expression
%%%%%%%%%%%%%%%%%%%%%%
\begin{eqnarray}
\Delta{\hat{k}_f} \ori = \ori - {\frac{1}{2 {\hat{s}}^2 Q_f}} \ori
( \; F_V^{(f)} \; - {\frac{v_f}{a_f}} \;  F_A^{(f)} )\;.
\end{eqnarray}
%%%%%%%%%%%%%%%%%
%The relations between vector and axial couplings are
%\begin{eqnarray}
%F^{(f)}_V &\equiv& u_L + u_R + \frac{F^{(f)}_L +F^{(f)}_R}{16\pi^2} \\
%F^{(f)}_A &\equiv& u_L - u_R + \frac{F^{(f)}_L -F^{(f)}_R}{16\pi^2} \\
%\end{eqnarray}
%%%%%%%%%%%%%%

In eq. (\ref{dkapaf}) we have denoted by $\delta k_f^{SUSY}$ the
supersymmetric contributions to $\Delta \hat{k}_f$.
Here we consider the example of the decoupling of large logs
in $\delta k_f^{SUSY}$ in
the case where the fermion $f$ stands for a ``down" quark denoted by $b$ 
being in the same
isospin multiplet with the ``up" quark denoted by $t$. In this case
we have
%%%%%%%%%%%%%%%%%%%
\begin{eqnarray}
u_L &=& -\frac{1}{2} + \frac{1}{3}\hs \;, \\ 
u_R &=& \frac{1}{3}\hs \;, \\
Q_f &=& -\frac{1}{3}\;.
\end{eqnarray}
%%%%%%%%%%%%%%%%%
The cases of other fermion species are treated in a similar manner.
In what follows we will consider only the chargino corrections to vertices
and external fermion lines. The decoupling of large logarithmic terms
arising from 
the neutralinos exchanges proceeds in exactly the same manner.

We will first discuss the self energy corrections to $Zb\bar{b}$
vertex. From the  diagrams of the Figure 2a, we obtain
{\footnote{ The functions $b_1, c_0$ used throughout this section
which are defined below should not be confused with the Passarino-Veltman
functions \cite{passarino} which are commonly denoted by capital letters.
These are
actually the reduced Passarino-Veltman functions \cite{ahn} defined as
%%%%%%%%%%%%%%%%%
\begin{eqnarray} 
b_1(m_1,m_2,q) &\equiv& \int_{0}^{1} dx\; x \log 
\frac{x m_1^2+(1-x) m_2^2-q^2 x (1-x) -i \epsilon}{Q^2}\;, \nonumber \\[1mm] 
c_0(m_1,m_2,m_3) &\equiv& \int_0^1 dx
\int_0^{1-x}dy \log \frac{(1-x-y) m_1^2+x m_2^2+y m_3^2-
(1-x-y)(x+y) m_b^2-x y P^2}{Q^2}\;. \nonumber
\end{eqnarray}
}}
, in an obvious notation,
%%%%%%%%%%%%%%%%
\begin{eqnarray}
F_L^{(b)} &=& \sum_{i=1,2}\sum_{j=1,2}\; b_1(m_{\ti{t}_j},
m_{\ti{\chi}_i},m_b)\;
u_L\;|a_{ij}^{b\ti{t}\ti{\chi}}|^2 \;, \\[2mm]
F_R^{(b)} &=& \sum_{i=1,2}\sum_{j=1,2}\; b_1(m_{\ti{t}_j},
m_{\ti{\chi}_i},m_b)\;
u_R\;|b_{ij}^{b\ti{t}\ti{\chi}}|^2 \;.
\end{eqnarray}
%%%%%%%%%%%%%%%%%%
 
%%%%%%%%%%%%%%%
%\vspace{.25in}
%\centerline{\hbox{\psfig{figure=fig1a.ps,width=2.5in}
%\psfig{figure=fig1b.ps,width=2.5in}}}
%\begin{center}
%\footnotesize{{\bf Figure 2 :} External fermion corrections to the 
%$Zb\bar{b}$ vertex.}
%\end{center}
%%%%%%%%%%%%%%%%%%%%%

>From the Appendix A (see the discussion following eqs. A.19) 
we get $a_{11}^{b\ti{t}\ti{\chi}}=g$,
$a_{22}^{b\ti{t}\ti{\chi}}=-h_t$ and $b_{21}^{b\ti{t}\ti{\chi}}=-h_b$. 
All other couplings vanish when the electroweak symmetry breaking effects are
ignored. Thus we get,
%%%%%%%%%%%%%%%%
\begin{eqnarray}
F_L^{(b)} &=& u_L \biggl [\; g^2 b_1(m_{\ti{t}_L},M_2,m_b)+
h_t^2 b_1(m_{\ti{t}_R},\mu,m_b)\;\biggr ]\;, \\[2mm]
F_R^{(b)} &=& u_R \biggl [\; h_b^2 b_1(m_{\ti{t}_L},\mu,m_b)\;\biggr ]\;.
\end{eqnarray}
%%%%%%%%%%%%%%%%

%\subsubsection{ Vertex corrections to $Zb\bar{b}$ vertex }

On the other hand, from the first triangle graph of Figure 2b we obtain,
%{\footnote{ $c_0(m_1,m_2,m_3)=\int_0^1 dx
%\int_0^{1-x}dy \log (\frac{(1-x-y) m_1^2+x m_2^2+y m_3^2-
%(1-x-y)(x+y) m_b^2-x y P^2}{Q^2})$}}

%%%%%%%%%%%%%%%
%\vspace{.25in}
%\centerline{\psfig{figure=fig2.ps,width=3in}}
%\begin{center}
%\footnotesize{{\bf Figure 3 :} Squark and chargino contribution to
%the $Zb\bar{b}$ vertex.}
%\end{center}
%%%%%%%%%%%%%%%%%%%%%

%%%%%%%%%%%%%%%%
\begin{eqnarray}
F_L^{(b)} &=& \sum_{i,j,k=1,2}\;
c_0(m_{\ti{\chi}_k},m_{\ti{t}_i},m_{\ti{t}_j})\;
\left (\frac{2}{3}\hs\delta_{ij}-\frac{1}{2} K_{i1}^{\ti{t} *}
K_{j1}^{\ti{t}}\right )\;a_{ki}^{b\ti{t}\ti{\chi}}
a_{kj}^{b\ti{t}\ti{\chi} *}\;, \\[2mm]
F_R^{(b)} &=& \sum_{i,j,k=1,2}\;
c_0(m_{\ti{\chi}_k},m_{\ti{t}_i},m_{\ti{t}_j})\;
\left (\frac{2}{3}\hs\delta_{ij}-\frac{1}{2} K_{i1}^{\ti{t} *}
K_{j1}^{\ti{t}}\right )\;b_{ki}^{b\ti{t}\ti{\chi}}
b_{kj}^{b\ti{t}\ti{\chi} *}\;,
\end{eqnarray}
%%%%%%%%%%%%%%%%%%
which, when the electroweak effects are ignored, have the following form,  
%%%%%%%%%%%%%%%%
\begin{eqnarray}
F_L^{(b)} &=& \left (\frac{2}{3}\hs-\frac{1}{2}\right )\, g^2
c_0(M_2,m_{\ti{t}_L},m_{\ti{t}_L})+\frac{2}{3}\,\hs h_t^2
c_0(\mu,m_{\ti{t}_R},m_{\ti{t}_R})\;,\\[2mm]
F_R^{(b)} &=&\left (\frac{2}{3}\hs-\frac{1}{2}\right )\, h_b^2
c_0(\mu,m_{\ti{t}_L},m_{\ti{t}_L})\;.
\end{eqnarray}
%%%%%%%%%%%%%%%%%
 
The calculation of the second diagram of Figure 2b gives
{\footnote{
%\begin{eqnarray}
$[c_2,c_6](m_1,m_2,m_3)=\int_0^1 dx\int_0^{1-x}[1,x]\;
\frac{1}{(1-x-y) m_1^2+x m_2^2+y m_3^2-(1-x-y) (x+y) m_b^2
-x y P^2-i\epsilon}\;.$
%\end{eqnarray}
}}

%%%%%%%%%%%%%%%%%%%%
%\vspace{.25in}
%\centerline{\psfig{figure=fig3.ps,width=3in}}
%\begin{center}
%\footnotesize{{\bf Figure 4 :} Chargino and squark contribution to
%the $Zb\bar{b}$ vertex.}
%\end{center}
%%%%%%%%%%%%%%%%%
 
%%%%%%%%%%%%%%%%
\begin{eqnarray}
F_L^{(b)} &=&-\sum_{i,j,k=1,2}\;
\Biggl \{\;\biggl [{P^2} c_6(m_{\ti{t}_k},m_{\ti{\chi}_i},
m_{\ti{\chi}_j})-\frac{1}{2}-c_0(m_{\ti{t}_k},m_{\ti{\chi}_i},
m_{\ti{\chi}_j})\;\biggr ]{\cal A^L}_{ij}\nonumber \\[2mm] &+&
{m_{\ti{\chi}_i} m_{\ti{\chi}_j}}\;
c_2(m_{\ti{t}_k},m_{\ti{\chi}_i},m_{\ti{\chi}_j})\; {\cal A^R}_{ij}\Biggr \}
a_{ik}^{b\ti{t}\ti{\chi}}a_{jk}^{b\ti{t}\ti{\chi} *}\;, \\[2mm]
%%%%%%%%%%%%%%%
F_R^{(b)} &=&-\sum_{i,j,k=1,2}\;
\Biggl \{\;\biggl [{P^2} c_6(m_{\ti{t}_k},m_{\ti{\chi}_i},
m_{\ti{\chi}_j})-\frac{1}{2}-c_0(m_{\ti{t}_k},m_{\ti{\chi}_i},
m_{\ti{\chi}_j})\;\biggr ]{\cal A^R}_{ij}\nonumber \\[2mm] &+&
{m_{\ti{\chi}_i} m_{\ti{\chi}_j}}\;
c_2(m_{\ti{t}_k},m_{\ti{\chi}_i},m_{\ti{\chi}_j})\; {\cal A^L}_{ij}\Biggr \}
b_{ik}^{b\ti{t}\ti{\chi}}b_{jk}^{b\ti{t}\ti{\chi} *}\;, 
\end{eqnarray}
%%%%%%%%%%%%%%%%%
where $P$ is the momentum carried by the Z - boson. 
The couplings ${{\cal A}^L}_{ij},{{\cal A}^R}_{ij}$ can be read from
Appendix A (see Eqs. A.17). In the absence of electroweak symmetry breaking
effects the only non-vanishing couplings are
%%%%%%%%%%%
\begin{eqnarray}
{{\cal A}^L}_{11} &=& \hc={{\cal A}^R}_{11}\;,\\[2mm]
{{\cal A}^L}_{22} &=& \hc-\frac{1}{2}={{\cal A}^R}_{22}\;.
\end{eqnarray}
%%%%%%%%%%%%%%%%
Thus, we obtain
%%%%%%%%%%%%%

%%%%%%%%%%%%%
\begin{eqnarray}
F_L^{(b)} &=& g^2 \hc c_0(m_{\ti{t}_L},M_2,M_2)+
h_t^2 \; \left (\hc-\frac{1}{2}\right )\; c_0(m_{\ti{t}_R},\mu,\mu)\;,\\[2mm]
F_R^{(b)} &=& h_b^2\;\left (\hc-\frac{1}{2}\right )\;c_0(m_{\ti{t}_L},\mu,\mu)
\;.
\end{eqnarray}
%%%%%%%%%%%%%%%%%%%

%\subsubsection{ The Decoupling of Large Logs }
 
Summing up the diagrams of  Figures 2a and 2b we get
%%%%%%%%%%%
\begin{eqnarray}
F_L^{(b)} &=& \left (-\frac{1}{2}+\frac{1}{3} \hs \right)
\left [\; g^2 b_1(m_{\ti{t}_L},M_2,m_b)+
h_t^2 b_1(m_{\ti{t}_R},\mu,m_b)\;\right ] \nonumber \\[2mm] &+&
\left (\frac{2}{3}\hs-\frac{1}{2}\right ) g^2
c_0(M_2,m_{\ti{t}_L},m_{\ti{t}_L})+\frac{2}{3}\hs h_t^2
c_0(\mu,m_{\ti{t}_R},m_{\ti{t}_R}) \nonumber \\[2mm] &+&
g^2 \hc c_0(m_{\ti{t}_L},M_2,M_2)+
h_t^2 \; \left (\hc-\frac{1}{2}\right )\; c_0(m_{\ti{t}_R},\mu,\mu)\;,\\[4mm]
%%%%%%%%%%%%%%%%%%%%%%%%%
F_R^{(b)} &=& \frac{1}{3}\hs
  h_b^2 b_1(m_{\ti{t}_L},\mu,m_b)\;
+
\left (\frac{2}{3}\hs-\frac{1}{2}\right )h_b^2\,
c_0(\mu,m_{\ti{t}_L},m_{\ti{t}_L})
\nonumber \\[2mm] &+&
h_b^2\;\left (\hc-\frac{1}{2}\right )\;c_0(m_{\ti{t}_L},\mu,\mu)\;.
\end{eqnarray}
%%%%%%%%%%%%%%%%%%%
 
In the limit of $q^2=m_b^2\simeq 0$, $P^2= M_Z^2\rightarrow 0$
or $M_Z^2<<M_{SUSY}^2$, the following useful relations hold,
%%%%%%%%%%%
\begin{eqnarray}
c_0(m_1,m_2,m_3)&=&b_1(m_2,m_1,0)\;, \\[2mm]
c_0(m_1,m_2,m_3)-b_1(m_1,m_2,0)&=&\frac{1}{m_1^2-m_2^2} \;
\left [m_1^2 m_2^2 \log\left (\frac{m_1^2}{m_2^2}\right )
-\frac{1}{2}\left (m_1^2+m_2^2\right )\;\right ]\;.
\end{eqnarray}
%%%%%%%%%%%%%%%%%%%

Using these we have for the expressions for $F^{(b)}_{L,R}$ above 
%%%%%%%%%%%%%
\begin{equation}
F_L^{(b)}= h_t^2\; O\left (\frac{m_{\ti{t}_R}^2}{\mu^2}\right )+
g^2\; O\left (\frac{m_{\ti{t}_L}^2}{M_2^2}\right )\;,
\end{equation}
%%%%%%%%%%%%%%%%
and
%%%%%%%%%%%
%%%%%%%%%%%%%
\begin{equation}
F_R^{(b)}=h_b^2 \; O\left (\frac{m_{\ti{t}_L}^2}{\mu^2}\right )\;,
\end{equation}
%%%%%%%%%%%%%%
which is independent of large logs and the decoupling of
terms $\log(\frac{M_{SUSY}}{M_Z})$ is manifest.

So far we have considered the cancellation of potentially large logarithms
involving the soft SUSY breaking scale $M_{1/2}$ and the mixing parameter
$\mu$ which arise from the neutralino and chargino sectors when
$M_{1/2}>>M_Z$. A similar analysis can be repeated for the corresponding
contributions of the squark and slepton sectors, whose masses depend also on
the soft SUSY breaking parameters $M_0$, when $M_0$ gets large.
We have carried out
such an analysis and found that the decoupling of large logarithms does indeed
occur when these parameters get large values. It is not necessary to
 present the details
of such a calculation here. We merely state that large logarithms arising
from the vector boson self energy corrections which contribute to the
quantities
$\Delta \hat r_W$, $\hat\rho$ and $\Delta \hat k_f$ cancel against those
from ${\hat{\alpha}}(M_Z)$. Also, the large log contributions 
from the muon decay amplitude,
which affect the effective mixing angle through $\delta_{VB}^{SUSY}$, 
cancel among themselves.
As for the large logarithmic contributions to the weak mixing angle 
from the non-universal corrections to the factor
$\Delta \hat k_f$, these are found to be cancelled in exactly the same
way as in the case of the neutralinos and charginos\footnote{The
logarithmic corrections of the Higgs sector to the $Zb\bar{b}$
vertex and external $b$ lines are cancelled in exactly the same 
manner.}.

%%%%%%%%%%%%%%%%%%%%%%%%%%%%%%%%%%%%%%%%%%%%%%%%%%%%%%%%%%%%%%%%%%%%%%%
%%%%%%%%%%%%    SQCD part %%%%%%%%%%%%%%%%%%%%%%%%%%%%%%%%%%%%%%%%%%%%%
%%%%%%%%%%%%%%%%%%%%%%%%%%%%%%%%%%%%%%%%%%%%%%%%%%%%%%%%%%%%%%%%%%%%%%%%
\vspace*{1cm}
{\bf{iv) \underline{SQCD corrections to $\Delta\hat{k}_f$}}}

\vspace*{.4cm} 
 The last corrections to be considered are the SQCD non-universal corrections
\cite{sqcdrefs}
which, due the largeness of the strong coupling constant, are,
naively, expected to
yield contributions larger than those of the electroweak sector. This case
is of relevance only when the external fermions in the $Zf\overline f$ vertex
are quark fields and is of particular interest for the bottom case whose
measurement of the Forward / Backward asymmetry ${\cal A}_b^{FB}$ 
yields the most precise individual measurement at LEP.

The one loop correction to $Zq\overline q$ vertex ({\it see} Figure 2c)
where two squarks, which
are coupled to the Z - boson, and a gluino are exchanged yields for the
Left and Right handed couplings defined in Eqs. (30)-(33),
%%%%%%%%%%%%%%%%
\begin{eqnarray}
F_L^{(q)} &=& {\frac{16}{3}} \; (4 \pi {\alpha}_s ) \;
\sum_{i=1,2}\sum_{j=1,2}\;
{K^{\tilde q}}^{\star}_{j1} K^{\tilde q}_{i1}  \;  A_{\tilde q}^{ji} \;
{C_{24}}(m_q^2,M_Z^2,m_q^2; M_{\tilde g}^2,m_{\tilde {q_i}}^2,
m_{\tilde {q_j}}^2) \;, \\[2mm]
F_R^{(q)} &=& {\frac{16}{3}} \; (4 \pi {\alpha}_s ) \;
\sum_{i=1,2}\sum_{j=1,2}\;
{K^{\tilde q}}^{\star}_{j2} K^{\tilde q}_{i2}  \;  A_{\tilde q}^{ji} \;
{C_{24}}(m_q^2,M_Z^2,m_q^2; M_{\tilde g}^2,m_{\tilde {q_i}}^2,
m_{\tilde {q_j}}^2)  \;.
\end{eqnarray}
%%%%%%%%%%%%%%%%%%
In these, the coupling $ A_{\tilde q}^{ji}$ is given by
%%%%%%%%%%%%%%%%
\begin{eqnarray}
A_{\tilde q}^{ji} \; = \;
u_L \; {K^{\tilde q}}^{\star}_{j1} K^{\tilde q}_{i1} +
u_R \; {K^{\tilde q}}^{\star}_{j2} K^{\tilde q}_{i2}\;,  \nonumber
\end{eqnarray}
%%%%%%%%%%%%%%%%%%
with $K^{\tilde q}_{ij}$ the matrix diagonalizing the squark ${\tilde q} $
mass matrix. The function $C_{24}$, with momenta and masses as shown,
is the coefficient of  $g_{\mu \nu}$ in the tensor three point integral
( This is denoted by $C_{20}$ in ref. \cite{Hollik2} ). The contribution
of ${F_{L,R}}^{(q)}$ to the form factor $\Delta\hat{k}_q$ is free of large
logarithms. In order to understand this consider the case of vanishing
quark mass $m_q$. In this case the matrix $K^{\tilde q}_{ij}$ becomes the
unit matrix.  It is easy to see that the contribution to
$\Delta\hat{k}_q$, as this is read from Eq. (36), is proportional to the
difference
\begin{eqnarray}
{C_{24}}(m_q^2,M_Z^2,m_q^2; M_{\tilde g}^2,m_{\tilde {q_1}}^2,
m_{\tilde {q_1}}^2)  - 
{C_{24}}(m_q^2,M_Z^2,m_q^2; M_{\tilde g}^2,m_{\tilde {q_2}}^2,
m_{\tilde {q_2}}^2) \;.  \nonumber
\end{eqnarray}
%%%%%%%%%%%%%%%%%%
In this difference the leading log terms cancel each other. Note that
 it would 
vanish if the left and right handed squark fields happened to
be degenerate in mass. Due to their mass splitting however
the result is not vanishing but at any rate small.
In general the SQCD vertex
corrections turn out to be smaller than the corresponding
electroweak corrections, as we have verified numerically.

 As for the external quark contributions ({\it see} Figure 2d) we find 
%%%%%%%%%%%%%%%%
\begin{eqnarray}
F_L^{(q)} &=& {\frac{8}{3}} \; (4 \pi {\alpha}_s ) \; u_L \;
\left [ c^2 \; {B_1}(m_q^2,M_{\tilde g}^2,m_{\tilde {q_1}}^2) +
  s^2 \; {B_1}(m_q^2,M_{\tilde g}^2,m_{\tilde {q_2}}^2)\right ]\;, \\
F_R^{(q)} &=& {\frac{8}{3}} \; (4 \pi {\alpha}_s ) \; u_R \;
\left [ s^2 \; {B_1}(m_q^2,M_{\tilde g}^2,m_{\tilde {q_1}}^2) +
  c^2 \; {B_1}(m_q^2,M_{\tilde g}^2,m_{\tilde {q_2}}^2)\right ] \;.
\end{eqnarray}
%%%%%%%%%%%%%%%%%%
In the equation above
$c \equiv K^{\tilde q}_{11}, s \equiv K^{\tilde q}_{12} $. Their contribution
to $\Delta\hat{k}_q$ is free of large logarithms and
small due to cancellations of the leading terms exactly as
in the case of the vertex corrections discussed previously. In fact in the
limit of vanishing quark mass the self energy corrections to 
$\Delta\hat{k}_q$ is proportional to the difference
\begin{eqnarray}
{B_1}(m_q^2,M_{\tilde g}^2,m_{\tilde {q_1}}^2) -
 {B_1}(m_q^2,M_{\tilde g}^2,m_{\tilde {q_2}}^2) \;. \nonumber
\end{eqnarray}
%%%%%%%%%%%%%%%%%%
which vanishes when the squark masses are equal. Therefore,
following the same arguments as in the vertex case, we are led to the
conclusion that SQCD contributions from the 
external quark lines are small. 

Besides the cancellations discussed above which lead to relatively small
SQCD vertex and external fermion corrections, these two contributions 
tend to cancel each other since they contribute with opposite signs. This
results in very small overall SQCD corrections to $\Delta\hat{k}_q$
almost one to two orders of magnitude smaller than the
corresponding non-universal electroweak corrections. We shall come back to
this point later when discussing our numerical results.

In the following section we shall discuss our numerical results
concerning the predictions of the MSSM for the effective mixing angle and
asymmetries. We will also
present the corresponding theoretical predictions for the mass
of the W - boson through its connection to the parameter rho and the
effective mixing angle.

\section{Numerical Analysis and Results}

For a given set of pole masses $m_t^{pole}$, $m_b^{pole}$, $m_{\tau}^{pole}$
we define the $\overline{DR}$ Yukawa couplings at $M_Z$.
To start with, we set a test value
for the $\hs$ ({\it i.e} $\hs=0.2315$)  and we define the 
$\overline{DR}$ gauge couplings $\hat{g}_1$ and $\hat{g}_2$ at $M_Z$. The
numerical output is independent of the starting value for $\hs$. For $\hs$ 
around the value given above the number of iterations needed for convergence
is minimized.
Then we use the 2-loop Renormalization Group equations 
\cite{2loop} to run up to
the scale $M_{GUT}$ where $\hat{g}_1$ and $\hat{g}_2$ meet. At 
$M_{GUT}$ we impose the unification condition
%%%%%%%%%%%%%%%
\begin{equation}
g_{GUT} \equiv \hat{g}_1 = \hat{g}_2 = \hat{g}_3 \;.
\end{equation}
%%%%%%%%%%%%%%%
Assuming universal boundary conditions for the soft breaking 
parameters $M_0$, $M_{1/2}$ and $A_0$, we run down to $M_Z$ and find the
couplings and 
the soft masses at $M_Z$ which are inputs for the self energies of the
gauge bosons, wave functions and vertex corrections and they
define the new $\hs$. The whole procedure is iterated until
convergence is reached satisfying the full one loop minimization
conditions in order to have radiative symmetry breaking observing the 
experimental bounds on supersymmetric particles.
For the calculation of the one loop integrals encountered we have made use of
the {\tt FF} library  \cite{olden}.
The conversion
of the ``theoretical'' $\hs$ to the experimental $s_f^2$ through
eq.(\ref{sineff})  gives our basic output : the effective weak
mixing angle $s_f^2$. In addition, the value of the strong QCD
coupling, as it is calculated in the $\overline {MS}$ scheme at $M_Z$, is among
our outputs \cite{dedes2}. Note that we
have used as inputs the parameters  $\alpha_{EM}$, $M_Z$ 
and $G_F$ which are experimentally known to a high degree of accuracy,
as well the masses of leptons and quarks.

%%%%%%%%%%%%%%%%%%%%%%%%%%%%%%%%%%%%%%%%%%%%%%%%%%%%%%%%%%%%%%%
%%%%%%%% DISCUSSION ON DK's%%%%%%%%%%%%%%%%%%%%%%%%%%%%%%%%%%%%
%%%%%%%%%%%%%%%%%%%%%%%%%%%%%%%%%%%%%%%%%%%%%%%%%%%%%%%%%%%%%%%%
The factor $\Delta \hat k_f$ needed to pass from the $\hat{s}^2(M_Z)$ to
the effective angle $s_f^2(M_Z)$ receives universal corrections, from the
${\gamma} - Z$ propagator, and non-universal corrections arising from
vertices and external wave function renormalizations. We find that
the non-universal Electroweak supersymmetric corrections are very small.
Although separately vertex and external fermion corrections are large they
cancel each other yielding contributions almost two orders of magnitude
smaller than the rest of the electroweak corrections. The non-universal
$SQCD$ contributions although a priori expected to to be larger than the
Electroweak corrections turn out to be even smaller. The reason for this was
explained in the previous section. In fact they are found to be one to two
orders of magnitude smaller than the corresponding electroweak corrections.
We conclude therefore that at the present level of accuracy one can
safely ignore the supersymmetric non-universal corrections to the factors
$\Delta \hat k_f$. The situation is very clearly depicted in Table I where
for some characteristic input values we give the contributions of the various
sectors to $\Delta \hat k_f$, as well as their total contributions, and also
the corresponding predictions for the 
values of the effective mixing angle and the asymmetries. Concerning the
values displayed in Table I, in a representative case, a few
additional remarks are in order: \\
i) The bulk of the supersymmetric corrections to $\Delta \hat k_f$
is carried by the universal corrections which are sizable, due to
their dependence on large logarithmic terms. These cancel similar terms in
$\hat{s}^2$. \\
ii) The contribution of Higgses, which is small, mimics that of the Standard
Model with a mass in the vicinity of $\simeq 100 GeV$. \\
iii) Gauge and Higgs boson contributions tend to cancel large universal
contributions of matter fermions. Concerning the gauge boson contributions
note that they are different for the different fermion species $l,c,b$.
This is due to the fact that their non-universal corrections depend on the
charge and weak isospin assignments of the external fermions and on
the mass of the top for when the external fermion is a bottom.  \\
iv) The slepton universal corrections are suppressed relative to their
corresponding squark contributions. This is due to the following reason.
The couplings of the left and right handed sleptons to the neutral
Z - boson depend on the angle $\hat{s}^2$ and would be exactly opposite if
$\hat{s}^2$ happened to be $\frac{1}{4}$. Thus their 
contributions to the
${\gamma} - Z$ propagator would be exactly opposite if their
masses were equal leading to a vanishing slepton contribution.The fact that
$\hat{s}^2 \simeq .23$ is close to $\frac{1}{4}$ in conjunction with the fact
that the left and right handed sleptons are characterized by small mass
splittings leads to the conclusion that universal slepton contributions to
$\Delta \hat k_f$ are small.
%%%%%%%%%%%%%%%%%%%%%%%%%%%%%%%%%%%%%%%%%%%%%%%%%%%%%%%%%%%%%%%%%%%%%%%
%%%%%%%%%%%% END of discussion on DK's%%%%%%%%%%%%%%%%%%%%%%%%%%%%%%%%%%
%%%%%%%%%%%%%%%%%%%%%%%%%%%%%%%%%%%%%%%%%%%%%%%%%%%%%%%%%%%%%%%%%%%%%%%%

In Figure 3, we display the effective weak mixing angle $s_l^2(M_Z)$,
 obtained from
the vertex $Z-l^+ - l^-$,  and the weak mixing
angle $\hat{s}^2(M_Z)$  as  functions of the soft gaugino
mass parameter $M_{1/2}$, the soft  parameter $M_0$, $A_0$ as well as
the parameters  $\tan\beta$ and $m_t$ inside the region which is indicated
in the figure.
Non-universal supersymmetric vertex and external fermion corrections
have been taken into account.
%The experimental values extracted from LEP, $s_l^2(M_Z)=0.23199 \pm 0.00028$,
%and SLD, $s_l^2(M_Z)=0.23055 \pm 0.00041$ 
%\cite{altarelli}, are also indicated in
% Figure 3 for comparison.
As is well known there is a discrepancy between the LEP and SLD
 experimental values of $s^2_{eff}$.
 The LEP average $s^2_{eff}=0.23199 \pm 0.00028$
 differs   by $2.9\sigma$ from the SLD value $s^2_{eff}=0.23055 \pm 0.00041$
obtained from the single measurement of left-right asymmetry \cite{altarelli}.
The LEP+SLD average value is $s^2_{eff}=0.23152+0.00023$.
We observe that $\hat{s}^2(M_Z)$
 takes on the ``theoretical'' value $\hat{s}^2=0.2377$ for
$M_{1/2}=900 \, GeV$ and  becomes larger and larger due to the fact
that it contains large
logarithms.
Manifest cancellation of large logarithmic terms is obtained in the
extracted value of the effective weak mixing angle as we have analytically
demonstrated in the previous chapter.
%Exploring the whole parameter space by giving arbitrary
%values in the plane $M_0$-$M_{1/2}$, we
%found that the effective weak mixing angle achieves its experimental LEP
%value quoted  above in a large range of the input values on the soft 
%parameter space.
%%%%%%%%%%%%%%%%%%%%%%%%%%%%%%%%%%%%
%%%%%%%%%%%%%%%%%%%%%%%%%%%%%%%
In Figure 3,
the dispersion of the values of
$s^2_l(M_Z)$ in the lower region of $M_{1/2}$ is caused by the
presence of the finite parts of order 
${\cal O}\left (M_Z/M_{SUSY}\right)$ in the expression (\ref{sineff}),
which become very important when $M_{1/2}\sim M_0\sim M_Z$ (case which
is preferred by SLD data) and 
 contribute
positively to $\Delta k$.
When $M_{1/2}\rightarrow 900 \; GeV$
(case which is rather preferred by LEP data) then $s^2_l(M_Z)\rightarrow
0.23145$ independently of the value of $M_0$. It must be noted that,
when $M_{1/2}=M_Z$ and $M_0\simeq 200\,GeV$,
 the values of the two angles are equal, {\it i.e.}
$s_l^2(M_Z)=\hat{s}^2(M_Z)=0.2310$.  We have also explored  the case
where the Higgs mixing parameter is negative ($\mu < 0$). In this case,
 as $M_{1/2}$ tends to larger
values $M_{1/2}\rightarrow 900 \, GeV$,   $s_l^2(M_Z)$
approaches 
the value $0.23145$ which means that, for large values of $M_{1/2}$, 
$s_l^2(M_Z)$ is independent of the sign of $\mu$ as it is expected
from the decoupling shown in chapter III. 
The sign of $\mu$ does not affect either  the $\hat{s}^2(M_Z)$ value
for large $M_{1/2}$.
 The effect of the sign
appears in the lower values of $M_{1/2}$. In the region $M_{1/2}\rightarrow
M_Z$, the value of $\Delta k_l$ ($l=e,\,\mu,\,\tau$)
 is always negative in the case $\mu < 0$
 and thus $s^2_l < \hat{s}^2$.
There is no possibility of equality between the two angles 
in this case. 
The largest value of $s_l^2(M_Z)=0.2315$
($\hat{s}^2(M_Z)\rightarrow 0.238$)
 is reached when
$M_{1/2}\rightarrow 1200\, GeV$. Just above this value
 no radiative symmetry breaking occurs. 
The lowest value of $s_l^2(M_Z)=0.2305$
($\hat{s}^2(M_Z)=0.2302$), for $\mu >0$, and
$s_l^2(M_Z)=0.2309$ ($\hat{s}^2(M_Z)=0.2316$) for $\mu<0$,
 is bounded by the new experimental limit on the chargino
mass which is around $\sim 84-86\, GeV$ \cite{Tata}.

%As is well known there is a discrepancy between the LEP and SLD
% experimental values of $s^2_f$. The LEP average $s^2_f=0.23199 \pm 0.00028$
% differs   by $2.9\sigma$ from the SLD value $s^2_f=0.23055 \pm 0.00041$
%obtained from the single measurement of left-right asymmetry \cite{altarelli}.
%The LEP+SLD average value is $s^2_f=0.23152+0.00023$.
In Figure 4, we plot
the values of $s^2_f$ for the fermions $f=c,b$. 
In the large SUSY breaking  limit, where all superparticles
are quite massive ($M_{1/2}\rightarrow 900 \, GeV$),
we obtain for the central values  $s^2_b=0.2330$ and $s^2_c=0.2314$. 
In the light limit, $M_{1/2}\simeq
M_0 \simeq M_Z$, they  take on the values,
  $s^2_b=0.2298$ and $s^2_c=0.2308$.
The main effect in the extracted values of the effective angle $s_l^2$
 is coming
dominantly from
the variation of $M_{1/2}$ and secondly from $M_0$.
If $M_{1/2}$ is kept constant, the variation
of $M_0$ from 100 to 900 GeV, changes  $s^2_f$ by +0.0005. In addition, the 
effect of $A_0$ on the effective angle is negligible. The effect of the
independent parameter $\tan\beta$ is also negligible if it remains in the
region $\tan\beta \simeq 5-28$.
Large loop corrections to the $b$-Yukawa coupling (or to the bottom pole mass),
which are proportional to
the term $\mu \tan\beta$, 
affect  the
obtained values of $s^2_f$ in the large $\tan\beta$ region \cite{Dedes3}.

There is a strong correlation of the output value of the effective weak mixing
angle with the top quark mass as it is shown in Figure 5. In this
Figure, we have chosen two characteristic sets of input values  $A_0=M_0=
M_{1/2}=600 \, GeV$ and  $A_0=M_0=M_{1/2}=200 \, GeV$.  It is
clear that the first case is most preferable if one assumes the LEP+SLD
data, where $s_l^2=0.23152 \pm 0.00023$. The present combined
CDF/D{\O} \cite{top} result for $m_t=175 \pm 5 \, GeV$,  
is 
also compatible with the first case.
Radiatively corrected light Higgs boson masses are also shown in Figure 5.
 Figures 6 and 7 display  the 
range of predictions for the mass of the W-gauge boson in the MSSM. As
one can see, the W-mass is in  agreement with the presently 
experimentally observed value, $M_W=80.427 \pm 0.075$
($M_W=80.405 \pm 0.089$) GeV obtained from LEP (CDF,UA2,D{\O}) experiments
\cite{LEP} for rather low (high) values of $M_{1/2}$ 
 in the region of $m_t=175 \pm 5$ GeV.
Variation of $m_t$ equal to $+5 \, GeV$ leads to variation of $M_W$ equal to
$+0.032 \, GeV$ while the effect on  $s_l^2$ is $-0.00017$.

The {\it left-right} asymmetries are given by the effective Lagrangian 
(\ref{leff}) with
%%%%%%%%%%%%%%%%%%
\begin{equation}
A_{LR}^f \ =\ {\cal A}^f \ =\ \frac{2\, \upsilon_{eff}^f / a_{eff}^f}
{1+\left ( \upsilon_{eff}^f / a_{eff}^f \right )^2 } \;,
\label{aet}
\end{equation}
%%%%%%%%%%%%%%%%%%%%%%
where
%%%%%%%%%%%%%%%
\begin{eqnarray}
\upsilon_{eff}^f \ &=& \ T_3^f - 2\, s^2_f \, Q^f \;, \nonumber \\[2mm]
a_{eff}^f \ &=& \ T_3^f \;.
\end{eqnarray}
%%%%%%%%%%%%%%%%%%%%
As it is depicted in Figure 8, the MSSM prediction for ${\cal A}^e$ agrees
with the LEP+SLD average value (${\cal A}^e = 0.1505 \pm 0.0023$) when
both $M_{1/2}$ and $M_0$
take on values around $M_Z$. In the heavy limit (large $M_{1/2}$),
 the 
MSSM agrees with the LEP value ${\cal A}^e = 0.1461 \pm 0.0033$. 
Note that as $M_{1/2}\rightarrow 900 \, GeV$, the value of ${\cal A}^e$ tends
asymptotically (which means that large logarithms have been decoupled
from the expression (\ref{aet}) ) to the value $0.1476$ corresponding
to $s^2_l \simeq 0.23145$ ({\it see} also Figure 4).

In the results shown in Figures 3-8, we have not considered the constraint
resulting from the experimental value of $\alpha_s$.
In Figure 9, we have plotted the acceptable values
of the soft breaking parameters $M_{1/2}$ and $M_0$\footnote{
We examine the region where $A_0\,,\,M_0\,,\,M_{1/2}\lesssim 900$ GeV.},
which are compatible with 
the  LEP+SLD ($\alpha_s=0.119
\pm 0.004$,  $s^2_{eff}=0.23152 \pm 0.00023$) \cite{altarelli}
 and the CDF/D{\O}
($m_t=175 \pm 5 \, GeV$) \cite{top} data.
The trillinear
soft couplings as well as the parameter $\tan\beta(M_Z)$ are taken
arbitrarily in the region ($0-900 \, GeV$) and ($2-30$), respectively.
 As we observe 
from  Figure 9, MSSM with radiative EW breaking
is valid in the region $M_{1/2} \gtrsim 500 \, GeV$ and
$M_0 \gtrsim 70 \, GeV$ \footnote{
The requirement that
the LSP is neutral puts this bound on $M_0$.}.
In this region, the physical gluino mass is above
$1 \, TeV$, the LSP (one of the neutralinos)
 is $ \gtrsim 200 \, GeV$, the chargino masses are
$m_{\tilde{\chi}_{1,2}} \gtrsim 650\, ,\, 370 \, GeV$, the stop masses are
$m_{\tilde{t}_{1,2}} \gtrsim 1000\, ,\, 790 \, GeV$, the sbottom masses are
$m_{\tilde{b}_{1,2}} \gtrsim 1000\, ,\, 960 \, GeV$, the slepton masses are
$m_{\tilde{\tau}_{1,2}} \gtrsim 340\, ,\, 210 \, GeV$, the sneutrinos are
$m_{\tilde{\nu}} \gtrsim 330 , GeV$ 
and the radiative 1-loop corrected Higgs masses
are $M_h, M_{A,H,H\pm} \gtrsim 108 \,,\, 780 \, GeV$, respectively.
Thus, we conclude that
the recent LEP+SLD and CDF/D{\O} data analysis
 favours the MSSM with radiative symmetry breaking
only  in the heavy limit of the sparticle masses.

\section{Conclusions}

We have considered the supersymmetric one loop corrections to the
effective mixing angle $s_f^2$ which is experimentally determined in LEP and
SLD experiments from measurements of on resonance left/right and
forward/backward asymmetries. This effective angle differs from the
corresponding mixing angle $\hs$ defined as the ratio of couplings which is
useful
to test unification of couplings in unified schemes encompassing the
Standard Model. The difference of the two angles, while very small in
the Standard Model, is substantial in supersymmetric extensions of it 
due to large logarithmic $log(\frac{M^2_{SUSY}}{M_Z^2})$ dependences
of $\hs$. Thus, although $\hs$ is a useful theoretical tool to test
the unification of couplings, it is not the proper quantity to compare with
experimental data which have already reached a high degree of accuracy.
Therefore, the relation between the two definitions is of utmost importance
for phenomenological studies of supersymmetric extensions of the Standard
Model.

In this article we have calculated all corrections to the factor
$\Delta{k_f}$ relating the two angles $s_f^2$ and $\hs$ including the
non-universal corrections from vertices and external fermions.
While $\Delta{k_f}$ is plagued by large logarithms in the limit where 
the supersymmetry breaking scale is large, the effective weak mixing angle
does not suffer from such large logarithms. 
In fact, we have proven that there are no dangerous logarithmic corrections
$log(\frac{M^2_{1/2}}{M_Z^2})$ from the chargino/neutralino sector
to the effective weak mixing angle. The decoupling of large logarithms
involving the Higgsino mixing parameter $\mu$, which in the constrained MSSM
with radiative symmetry breaking,
is large, is obtained in the same manner.
The cancellation of potentially dangerous terms also holds for the
contributions of the squark and slepton sector. The
cancellation of the  $log(\frac{M^2_{SUSY}}{M_Z^2})$ terms in the
$\overline{DR}$ scheme had been shown only numerically in previous studies.

It must be noted that there are large logarithmic terms 
in the    ``non-oblique'' supersymmetric
wave function renormalization of external fermions and vertex corrections
of the vertex $Zf\overline{f}$. Nevertheless, we have analytically proven  
that they get decoupled from $\Delta{k_f}$ and, hence, from
the effective weak mixing angle itself. In addition to
the analytical results described in  chapter III,
we have also displayed representative numerical results in Table I in two
particular cases of the MSSM.

We have also presented analytically, the decoupling of the
large logarithmic terms from $s^2_f$ in the case of the non-universal
SQCD corrections. Besides the self-cancellations of
this terms from the relevant diagrams Fig.2c and Fig.2d, there are
additional cancellations from the summation of these diagrams due
to their opposite sign. 
We have found that these corrections are very small and 
could be safely ignored from the analysis in the present experimental 
accuracy.

We have further proceeded to a numerical study of the one loop corrected
effective mixing angle having as inputs the values of $\alpha_{EM}$, $M_Z$,
the Fermi coupling constant $G_F$ and the experimental values for the
fermion masses. Assuming coupling constant unification and radiative
breaking of the electroweak symmetry we have scanned the  
soft SUSY breaking parametric space and given theoretical predictions for the
value of the effective mixing angles, the value of the strong coupling
constant at $M_Z$ and the value of the W - boson mass as this is determined
from the parameter $\rho$ and the effective weak mixing angle.
We find that the large logarithmic corrections of the
form $log (\frac{M_{SUSY}^2}{M_Z^2})$ indeed get decoupled  from the 
extracted value of the effective weak mixing angle in the region of
large $M_{1/2}$ and $M_0$ (Figure 3) following our 
analytical calculations. The predicted MSSM values of the effective
angles are in agreement with the LEP+SLD data (Figure 4)
 as well as
with the new CDF/D{\O} \cite{top}
 results for the top mass $m_t=175 \pm 5 \, GeV$ 
(Figure 5)
 in the region where
all superparticles are quite massive. In this region, MSSM predicts 
values of the W-gauge boson mass which are
 in  agreement  with the new \cite{LEP}
CDF,UA2,{D\O} average value $80.405 \pm 0.089 \, GeV$ (Figures 6,7).
Large logarithms are also decoupled from the {\it left-right} asymmetry value
${\cal A}^e$. MSSM seems to prefer the experimental LEP value of 
${\cal A}^e$, rather than
the average value from LEP+SLD (Figure 8). Finally, values
of $M_{1/2}$ which are greater than $500 \, GeV$ are favoured by
the MSSM if one assumes the present LEP and CDF/D{\O} data for
$s_l^2$, $\alpha_s$ and $m_t$ (Figure 9).

\vspace*{0.1cm}
{\noindent\bf Note Added :}

\vspace{.2cm}
After submitting this article for publication we became aware of the paper
by P. Chankowski and S. Pokorski \cite{Pok3} where corrections to the
leptonic mixing angle and predictions for the W boson mass are presented.

\vspace*{0.1cm}
{\noindent\bf Acknowledgements} 

\vspace{.2cm}
The authors wish to thank Peggy Kouroumalou who collaborated in the early stages 
of this work. A.D. and K.T. acknowledge financial support from the 
research program $\Pi{\rm ENE}\Delta$-95
of the Greek Ministry of Science and Technology. A.B.L. and K. T. acknowledge
 support from
the TMR network ``Beyond the Standard Model", ERBFMRXCT-960090. 
A. B. L. acknowledges 
support from the Human Capital and Mobility program CHRX-CT93-0319.

\newpage
{\noindent\bf Appendix A: Quick reference to neutralino/chargino
and their interactions}
\vspace{.7cm}
\setcounter{equation}{0}
\renewcommand{\theequation}{A.\arabic{equation}}

In the $\tilde{B}$, $\tilde{W}^{(3)}$, 
$i \tilde{H}_{1}^0$, $i \tilde{H}_{2}^0$, basis 
the neutralino mass matrix is
%%%%%%%%%%%%%%%%%%%
\begin{equation}
{\cal M}_N \ =\ \left(\begin{array}{cccc} M_1 & 0 &
g^\prime \frac{\upsilon cos\beta}{2} &
-g^\prime \frac{\upsilon sin\beta}{2}
\\[1mm] 0 & M_2 & -g \frac{\upsilon cos\beta}{2} &
g \frac{\upsilon sin\beta}{2}
\\[1mm] g^\prime \frac{\upsilon cos\beta}{2}&
-g \frac{\upsilon cos\beta}{2} & 0 & -\mu \\[1mm]
-g^\prime \frac{\upsilon sin\beta}{2}
&g \frac{\upsilon sin\beta}{2} & -\mu & 0
\end{array} \right)\ .\label{mchi0}
\end{equation}
%%%%%%%%%%%%%%
The mass eigenstates (${\tilde \chi}_{1,2,3,4}^0$)
 of neutralino mass matrix ${\cal M}_N$ are
written as 
%%%%%%%%%%%%%%%%%%%
\begin{equation}
{\cal O} \,\,  \left ( \begin{array}{c} {\tilde \chi}_1^0 \\
{\tilde \chi}_2^0 \\ {\tilde \chi}_3^0 \\ {\tilde \chi}_4^0
\end{array} \right )
\ =\ \left ( \begin{array}{c} \tilde{B} \\ \tilde{W}^{(3)} \\
i \tilde{H}_{1}^0 \\ i \tilde{H}_{2}^0
\end{array} \right )\;.
\end{equation}
%%%%%%%%%%%%%%%%%%%
and 
%%%%%%%%%%%%
\begin{equation}
{\cal O}^T {\cal M}_N {\cal O}\ ={\rm Diag} \left (
m_{{\tilde \chi}^0_1},m_{{\tilde \chi}^0_2}, m_{{\tilde \chi}^0_3},
m_{{\tilde \chi}^0_4} \right ) \;,
\end{equation}
%%%%%%%%%%%%
where ${\cal O}$ is a real orthogonal matrix. Note that when 
electroweak breaking effects are ignored ${\cal O}$ can get
the form
%%%%%%%%%%%%%%%%%%%
\begin{equation}
{\cal O} \ =\ \left ( \begin{array}{cc} {\bf 1_2} & {\bf 0_2} \\[1mm]
{\bf 0_2} & \begin{array}{cc} \frac{1}{\sqrt{2}} & \frac{1}{\sqrt{2}} \\
-\frac{1}{\sqrt{2}} & \frac{1}{\sqrt{2}} \end{array} \end{array} 
\right ) \;.
\end{equation}
%%%%%%%%%%%%%%%%%

The chargino mass matrix can be obtained from the following
Lagrangian mass terms
%%%%%%%%%%%%
\begin{equation}
{\cal L}^{mass}_{charginos} \ =\ 
- \left ( \tilde{W}^- , {i \tilde{H}_{1}^-} \right ) {\cal M}_c 
\left (\begin{array}{c} \tilde{W}^+ \\ {i \tilde{H}_{2}^+} \end{array}
\right ) \, + \, (h.c) \;,
\end{equation}
%%%%%%%%
where we have defined $\tilde{W}^\pm \equiv \frac{ \tilde{W}^{(1)}
\mp i \tilde{W}^{(2)} }{\sqrt{2}}$ and 
%%%%%%%%%
\begin{equation}
{\cal M}_C \ =\ \left (\begin{array}{cc} M_2 &
 -g \frac{\upsilon sin\beta}{\sqrt{2}} \\[1mm]
- g \frac{\upsilon cos\beta}{\sqrt{2}} & \mu \end{array}
\right )\; .
\label{chmat}
\end{equation}
%%%%%%%%%%%%
Diagonalization of this matrix gives
%%%%%%%%
\begin{equation}
U {\cal M}_c V^\dagger \ =\ \left (\begin{array}{cc} m_{\tilde{\chi}_1}
& 0 \\[1mm]
0 & m_{\tilde{\chi}_2} \end{array} \right ) \; .
\end{equation}
%%%%%%%%%%%%%%%%%%%
Thus,
%%%%%%%%%%%%
\begin{equation}
{\cal L}^{mass}_{charginos} \ =\ -m_{\tilde{\chi}_1} 
\bar{\tilde{\chi}_1} \tilde{\chi}_1 -  m_{\tilde{\chi}_2}
\bar{\tilde{\chi}_2} \tilde{\chi}_2 \; .
\end{equation}
%%%%%%%%%%%%%%
The Dirac chargino states $\tilde{\chi}_{1,2}$ are given by
%%%%%%%%%%%%%%%
\begin{equation}
\tilde{\chi}_1 \equiv \left (\begin{array}{c} \lambda_1^+ \\
\bar{\lambda}_1^- \end{array} \right ) \,\, , \,\,
\tilde{\chi}_2 \equiv \left (\begin{array}{c} \lambda_2^+ \\
\bar{\lambda}_2^- \end{array} \right ) \;.
\end{equation}
%%%%%%%%%%%%%%%%%
The two component Weyl spinors $\lambda^\pm_{1,2}$ are related
to $\tilde{W}^\pm$, ${i \tilde{H}_{1}^-}$, ${i \tilde{H}_{2}^+}$ by
%%%%%%%%%%%%%%%
\begin{equation}
V \left ( \begin{array}{c} \tilde{W}^+ \\ {i \tilde{H}_{2}^+}
\end{array} \right ) \equiv \left (\begin{array}{c}
\lambda_1^+ \\ \lambda_2^+ \end{array} \right ) \,\, , \,\,
\left ( \tilde{W}^- \, , \, {i \tilde{H}_{1}^-} \right ) U^\dagger
 \equiv \left ( \lambda_1^- \, , \, \lambda_2^- \right ) \;.
\end{equation}
%%%%%%%%%%%%%%%%

The gauge interactions of charginos and neutralinos can
be read from the following Lagrangian\footnote{
In our notation $\hat{e} \equiv $electron's charge just
opposite to that used in ref. \cite{Hollik2}.}
%%%%%%%%%%%%%%%
\begin{equation}
{\cal L} \ =\ \hat{g} \left ( W^+_\mu J_-^\mu + W_\mu^- J_+^\mu 
\right ) + \hat{e} A_\mu J_{em}^\mu +\frac{\hat{e}}{\hat{s} \hat{c}}
Z_\mu J_Z^\mu \;.
\end{equation}
%%%%%%%%%%%%%%%%%
Also,
%%%%%%%%%%%%%
\begin{equation}
\left (\begin{array}{c} Z_\mu \\[1mm] A_\mu \end{array} \right )
\ =\ \left ( \begin{array}{cc} \hat{c} & \hat{s} \\[1mm]
-\hat{s} & \hat{c} \end{array} \right ) \, 
\left ( \begin{array}{c} W_\mu^{(3)} \\[1mm] B_\mu \end{array}
\right ) \;.
\end{equation}
%%%%%%%%%%%%%%%%%
The currents $J_+^\mu$, $J_{em}^\mu$ and $J_Z^\mu$ are given by
%%%%%%%%%%%%%%%%%
\begin{equation}
J_+^\mu \equiv \bar {{\tilde \chi}^0_a}  \gamma^\mu \left [
{\cal P}_L {\cal P}^L_{a i}+ 
{\cal P}_R {\cal P}^R_{a i}  \right ]
\tilde{\chi}_i \;\; a=1...4,\;\; i=1,2 \;,
\end{equation}
%%%%%%%%%%%%%%
where
${\cal P}_{L,R} = \frac{1 \mp \gamma_5 }{2}$ and
%%%%%%%%%%%%%%%
\begin{eqnarray}
& &{\cal P}^L_{a i}\equiv +\frac{1}{\sqrt{2}} {\cal O}_{4 a} V^{*}_{i 2}
- {\cal O}_{2 a} V^{*}_{i 1}\;,\nonumber \\[2mm]
& &{\cal P}^R_{a i}\equiv -\frac{1}{\sqrt{2}} {\cal O}_{3 a} U^{*}_{i 2}
- {\cal O}_{2 a} U^{*}_{i 1} \;.
\end{eqnarray}
%%%%%%%%%%%%%% 
The electromagnetic current $J_{em}^\mu$ is
%%%%%%%%%%%%
\begin{equation}
J_{em}^\mu \ =\ \bar{\tilde\chi}_1 \gamma^\mu \tilde{\chi}_1 +
\bar{\tilde\chi}_2 \gamma^\mu \tilde{\chi}_2 \;.
\end{equation}
%%%%%%%%%%%%%%%%%%%
Finally, the neutral current $J_Z^\mu$ can be read from
%%%%%%%%%%%%%%%
\begin{equation}
J^\mu_Z \equiv \bar  {{\tilde \chi}_i}  \gamma^\mu \left [
{\cal P}_L {\cal A}^L_{i j} + {\cal P}_R {\cal A}^R_{i j} \right ]
\tilde{\chi}_j + \frac{1}{2} \bar {{\tilde \chi}^0_a} \gamma^\mu \left [
{\cal P}_L {\cal B}^L_{a b} + {\cal P}_R {\cal B}^R_{a b} \right ]
{{\tilde \chi}^0_b} \;,
\end{equation}
%%%%%%%%%%%%%%%%%%%%%
with
%%%%%%%%%%%%%%
\begin{eqnarray}
{\cal A}^L_{i j} &=& \hc \delta_{i j} -\frac{1}{2} V_{i 2} V^{*}_{j 2}
\;,\nonumber \\
{\cal A}^R_{i j} &=& \hc \delta_{i j} -\frac{1}{2} U_{i 2} U^{*}_{j 2}
\;,\nonumber \\
{\cal B}^L_{a b} &=& \frac{1}{2} \left ( {\cal O}_{3 a} {\cal O}_{3 b} -
{\cal O}_{4 a} {\cal O}_{4 b} \right )
\;,\nonumber \\
{\cal B}^R_{a b} &=& - {\cal B}^L_{a b} \;.
\end{eqnarray}
%%%%%%%%%%%%%%%%%%
Note that since ${\cal B}^R_{a b} = - {\cal B}^L_{a b}$ the neutralino
contribution to $J_Z^\mu$ can be cast into the form
%%%%%%%%%%%%%%%%%%%
\begin{equation}
J_Z^\mu \ =\ -\frac{1}{2} {\cal B}^L_{a b} \left ( \bar{{\tilde \chi}^0_a}
\gamma^\mu \gamma^5 {{\tilde \chi}^0_b} \right ) \;.
\end{equation}
%%%%%%%%%%%%%

For the supersymmetric external fermion corrections we need know the
chargino and neutralino couplings to fermions and sfermions.
The relevant chargino couplings are given by the following
Lagrangian terms
%%%%%%%%%%%%%
\begin{equation}
{\cal L} = i \; {\bar{\tilde {\chi}}}_{i}^{c} \;
 ({\cal P}_L \, a^{ {f^{\prime}}   {\tilde f} }_{ij} +
{\cal P}_R \, b^{ {f^{\prime}}   {\tilde f} }_{ij}) \,  {f^\prime } \,
 {\tilde f}_{j}^{*}
\, + \,
i \; {\bar{\tilde {\chi}}}_{i} \;
 ({\cal P}_L \, a^{f {\tilde f}^{\prime} }_{ij} +
{\cal P}_R \, b^{f {\tilde f}^{\prime} }_{ij}) \,  {f} \,
 {\tilde f}_{j}^{\prime *} \, + \, (h.c) \;.
\end{equation}
In this, ${\chi}_{i} \; (i=1,2)$ are the positively charged charginos  and 
${\chi}_{i}^{c}$ the corresponding charge conjugate states having 
opposite charge. $f\, , \,{f}^{\prime}$ are ``up" and ``down"
fermions, quarks or leptons, while ${\tilde f}_i\, , \,{\tilde f}_i^{\prime}$
are the corresponding sfermion mass eigenstates.
The left and right-handed couplings appearing above are given by
%%%%%%%%%%%%%%%%%%%%%%
\begin{eqnarray}
a^{{f^{\prime}}{\tilde f} }_{ij} \ori &= \ori
g V_{i1}^{*} \, K^{\tilde f}_{j1} - h_{f} V_{i2}^{*}  K^{\tilde f}_{j2} \ori ,
\ori  &b^{ {f^{\prime}} {\tilde f} }_{ij} \ori =
 \ori -h_{f ^\prime} \,  U_{i2}^{*} K^{\tilde f}_{j1}  \;,  \nonumber  \\
a^{f {\tilde f}^{\prime} }_{ij} \ori &= \ori
g U_{i1} \, K^{{\tilde f}^{\prime}}_{j1} + h_{f^\prime} \, U_{i2}
  K^{{\tilde f}^{\prime}}_{j2} \ori , \ori
&b^{f {\tilde f}^{\prime} }_{ij} \ori = \ori
h_{f } \,  V_{i2} K^{{\tilde f}^{\prime}}_{j1} \;.  \nonumber
\end{eqnarray}
%%%%%%%%%%%%%%%%%%%%%%
In the equation above $h_{f } \, , \, h_{f ^\prime} $ are the Yukawa
couplings of the up and down fermions respectively. The matrices
$K^{\tilde{f},{\tilde f}^\prime} $ which diagonalize the sfermion mass matrices become the
unit matrices in the absence of left-right sfermion mixings.
For the electron and muon family the lepton masses are taken to be
vanishing in the case that mixings do not occur. In addition the 
right-handed couplings, are zero.  \newline 
The corresponding neutralino couplings are given by
%%%%%%%%%%%%%%%%%%%%%%%%%%%%%%%%%%%%
\begin{equation}
{\cal L} = i \; {\bar  {{\tilde \chi}^0_a}} \;
 ({\cal P}_L \, a^{ {f}   {\tilde f} }_{aj} +
{\cal P}_R \, b^{ {f}   {\tilde f} }_{aj}) \,  {f} \,
 {\tilde f}_{j}^{*}
\, + \,
i \; {\bar{{\tilde \chi}^0_a}} \;
 ({\cal P}_L \, a^{{f^{\prime}} {\tilde f}^{\prime} }_{aj} +
{\cal P}_R \, b^{{f^{\prime}} {\tilde f}^{\prime} }_{aj}) \, {f^{\prime}} \,
 {\tilde f}_{j}^{\prime *} \, + \, (h.c) \;.
\end{equation}
%%%%%%%%%%%%%%%%%%%%%%%
The left and right-handed couplings for the up fermions, sfermions
are given by
%%%%%%%%%%%%%%%%%%%%%%
\begin{eqnarray}
a^{{f}{\tilde f}}_{aj} \ori &= &\ori
{\sqrt{2}} \, ( g {T^{3}_f}{O_{2a}}+{g^{\prime}}{\frac{Y_f}{2}} \,{O_{1a}})
\, {K^{f}_{j1}} \ori + \ori h_{f} \, {O_{4a}}\,  {K^{f}_{j2}}  \ori , \ori
\nonumber \\
b^{{f}{\tilde f} }_{aj} \ori & = & \ori
{\sqrt{2}} \, (-{g^{\prime}}{\frac{Y_{f^c}}{2}} \,{O_{1a}}) \, {K^{f}_{j2}}
\ori - \ori h_{f} \, {O_{4a}}\,  {K^{f}_{j1}} \;,  \nonumber  
\end{eqnarray}
%%%%%%%%%%%%%%%%%%%%%%
while those for the down fermions and sfermions are given by
%%%%%%%%%%%%%%%%%%%%%%%%%%%%%
\begin{eqnarray}
a^{{\fp}{\tilde \fp}}_{aj} \ori &= &\ori
{\sqrt{2}} \,
( g {T^{3}_{\fp}}{O_{2a}}+{g^{\prime}}{\frac{Y_\fp}{2}} \,{O_{1a}})
\, {K^{\fp}_{j1}} \ori - \ori h_{\fp} \, {O_{3a}}\,  {K^{\fp}_{j2}}  \ori , \ori
\nonumber \\
b^{{\fp}{\tilde \fp} }_{aj} \ori & = & \ori
{\sqrt{2}} \, (-{g^{\prime}}{\frac{Y^{\prime}_{f^c}}{2}} \,{O_{1a}})
 \, {K^{\fp}_{j2}}
\ori + \ori h_{\fp} \, {O_{3a}}\,  {K^{\fp}_{j1}}\;.   \nonumber  
\end{eqnarray}

\vspace*{1cm}
%\newpage
{\noindent\bf Appendix B: Passarino - Veltman functions }
\vspace{.7cm}
\setcounter{equation}{0}
\renewcommand{\theequation}{B.\arabic{equation}}

%%%%%%%%%%%%%%%%%%
All functions appearing in the propagator corrections in 
the main text can be expressed in terms of the basic Passarino - Veltman
integrals $A_0, B_0$ in the following way
%%%%%%%%%%%%%%%%%%%
\begin{equation}
A_0(m)= m^2 (\frac{1}{\hat{\epsilon}} + 1 - \ln\frac{m^2}{Q^2} )\;,
\end{equation}
%%%%%%%%%%%%%%%%%
\begin{equation}
B_0(p,m_1,m_2)=\frac{1}{\hat{\epsilon}}
-\int_0^1 dx \ln\frac{(1-x)m_1^2+x m_2^2
-x (1-x)p^2-i\epsilon}{Q^2}\;,
\end{equation}
%%%%%%%%%%%%%%%%%
where $\frac{1}{\hat{\epsilon}}=\frac{1}{\epsilon}-\gamma_E+\ln 4\pi$.
This reduction can be done with the following identities (in what follows
we made use of these functions only)
%%%%%%%%%%%%%
\begin{eqnarray}
H(p,m_1,m_2)&=&4\tilde{B}_{22}(p,m_1,m_2)+
(p^2-m_1^2-m_2^2)B_0(p,m_1,m_2)\;,\\[3mm]
\tilde{B}_{22}(p,m,m)&=&-\frac{1}{12}p^2 B_0(p,m,m)-\frac{1}{18}p^2
+\frac{1}{3}\left [m^2+m^2 B_0(p,m,m)-A_0\right ]\;,\\[3mm]
\tilde{B}_{22}(0,m,m)&=&0\;,\\[3mm]
B_0[p,m,m]&\stackrel{m^2>>p^2}{\longrightarrow}&-\ln(m^2/Q^2)\;.
\end{eqnarray}
%%%%%%%%%%%%%%%%%%%%%%%
For the vertex and box corrections to muon decay we need the
functions $B_0, B_1, C_0$
at zero momenta. The following relations are useful in order to express
the contributions to $\delta^{SUSY}_{VB}$ in terms of the
masses of the particles in the loop

\begin{eqnarray}
B_0(0,m_1,m_2) \ =\ {1\over\hat\epsilon} + 1 + \ln\left(Q^2\over
m_2^2\right) + {m_1^2\over m_1^2-m_2^2} \ln\left(m_2^2\over
m_1^2\right)~,
%\nonumber
\end{eqnarray}
\begin{eqnarray}
B_1(0,m_1,m_2) \ =\ {1\over2}\biggl[{1\over\hat\epsilon} + 1 +
\ln\left(Q^2\over m_2^2\right) + \left({m_1^2\over m_1^2 -
m_2^2}\right)^2\ln\left({m_2^2\over m_1^2}\right) +
{1\over2}\left({m_1^2+m_2^2\over m_1^2-m_2^2}\right)\biggr]~,
%\nonumber
\end{eqnarray}
%%%%%%%%%%%%%%%%%%%%%%
\begin{eqnarray}
C_0( m_1,m_2,m_3)\ =\ - \int_0^1 dx \int_0^{1-x} dy\, \frac{1}
{m_1^2 x +m_2^2 y +m_3^2 (1-x-y)}\;,
%\nonumber
\end{eqnarray}
%%%%%%%%%%%%%
%%%%%%%%%%%%%%%%%%%%%%
\begin{eqnarray}
C_0(m_1,m_2,m_2)\ =\ \frac{1}{m_1^2 -m_2^2} + \frac{m_1^2}{\left ( m_1^2
- m_2^2 \right )^2} \ln \left ( \frac{m_2^2}{m_1^2} \right ) \;.
%\nonumber
\end{eqnarray}
%%%%%%%%%%%%%%%

%%%%%%%%%%%%%%%%%%

%%%%%%%%%%%%%%%%%%%%%%%%%% TABLE %%%%%%%%%%%%%%%%%%%%%%%%%%
\newpage
%\draft
\mediumtext
\begin{table}
\caption{Partial and total contributions to ${\Delta k}_f$, 
($f=lepton,charm,bottom$), for two sets of inputs shown at the top.
Also shown are the predictions for the effective weak mixing angles and the
asymmetries. In the first five rows we display the universal contributions
to $10^3 \; \times \;{\Delta k}$ of squarks ($\tilde q$),
sleptons($\tilde l$), Neutralinos and Charginos (${\tilde Z},{\tilde C}$),
ordinary fermions and Higgses %$\; \; \;  $
(The number shown in the middle below the
"charm" column refers to "lepton" and "bottom" as well). In the next five rows
we display the contributions of gauge bosons as well as the supersymmetric
$Electroweak \; (EW)$ and $SQCD$ vertex
and external fermion wave function renormalization corrections
to $10^3 \; \times \;{\Delta k}$. }
%%%%%%
\begin{tabular}{ccccccccc}
 &$M_0=200$&$M_{1/2}=200$&$A_0=200$& &
 &$M_0=400$&$M_{1/2}=400$&$A_0=500$ \\
 &$m_t=175$&$tanb=4$&$\mu > 0$&  &
 &$m_t=175$&$tanb=4$&$\mu > 0$  \\
\tableline
\\
 &$lepton$ &$charm$ &$bottom$& &
 &$lepton$ &$charm$ &$bottom$ \\   \\
\tableline
\\
${\tilde q}$& &-6.6206 & && & &-8.8188 & \\
${\tilde l}$& &-0.3123 & && & &-0.3791 & \\
${\tilde Z},{\tilde C}$& &-4.4892 & && & &-9.6138 & \\
$Fermions$& &4.6573 & && & &4.5103 & \\
$Higgs$& &-0.7312 & && & &-1.0113 & \\  \\
$Gauge$&-3.1782  &-3.6272 &2.2911  &&
       &-3.1128   &-3.5572 &2.2822  \\ \\
$Vertex(EW)$&1.2324&3.5581 &12.7641  &&
       &2.2825   &4.8539 &18.1519  \\
$Wave(EW)$&-1.3209&-3.5208 &-12.9598 &&
       &-2.2973   &-4.8422 &-18.0059\\   \\
$Vertex(SQCD)$& - &0.2110 &-1.1129  &&
       & -  &0.2361 &-1.1166 \\
$Wave(SQCD)$&- &-0.2103 &1.1012 &&
       & -  &-0.2359 &1.1135\\
\tableline
\\ 
%$Total$ & & & && & & & \\
${\Delta k}\;(\times \; 10^2 ) $ &-1.0763 &-1.1085 &-0.5412 &&
&-1.8440 &-1.8858 &-1.2888  \\
%\tableline
${sin^2} {\theta}_f$& 0.23134 & 0.23126 &0.23259 && &0.23145
& 0.23135 & 0.23276\\
${\cal A}_{LR}^f$& 0.1485 & 0.6684 &0.9348 && &0.1476 &0.6681 & 0.9347\\
${\cal A}_{FB}^f$& 0.0165 & 0.0744 &0.1041 && &0.0163 & 0.0740 & 0.1035\\

\end{tabular}
\end{table}
%%%%%%%%%%%%%%%%%%%%%% End of TABLE I%%%%%%%%%%%%%%%%%%%%%%%%%%%%%%%%%

%%%%%%%%%%%%%%%%%%%%%   FIGURES %%%%%%%%%%%%%%%%%%%%%%%
%\newpage

\vspace*{1.5in}
%%%%%%%%%%%%%%%%%%%%%%%%%%
\centerline{\input{sinhat.ps}}
\begin{center}
\footnotesize{{\bf Figure 1:}
 The values of the running weak mixing angle $\hat{s}^2$ at
$M_Z$ in the $\overline{DR}$ scheme 
defined as a ratio of gauge couplings for various
input universal soft gaugino masses $M_{1/2}$
 for particular input of $M_0$ , $A$ , $\tan \beta$ and $m_{t}$ . 
 The strong dependence of $\hat{s}^2$ on $M_{1/2}$ near $M_Z$ is due to 
 the presence of sparticle thresholds.}
\end{center}
%%%%%%%%%%%%%%%%%%%%%%%%%%%

\newpage
%%%%%%%%%%%%%%%

%%%%%%%%%%%%%%%

%\vspace*{1in}

\centerline{\hbox{\psfig{figure=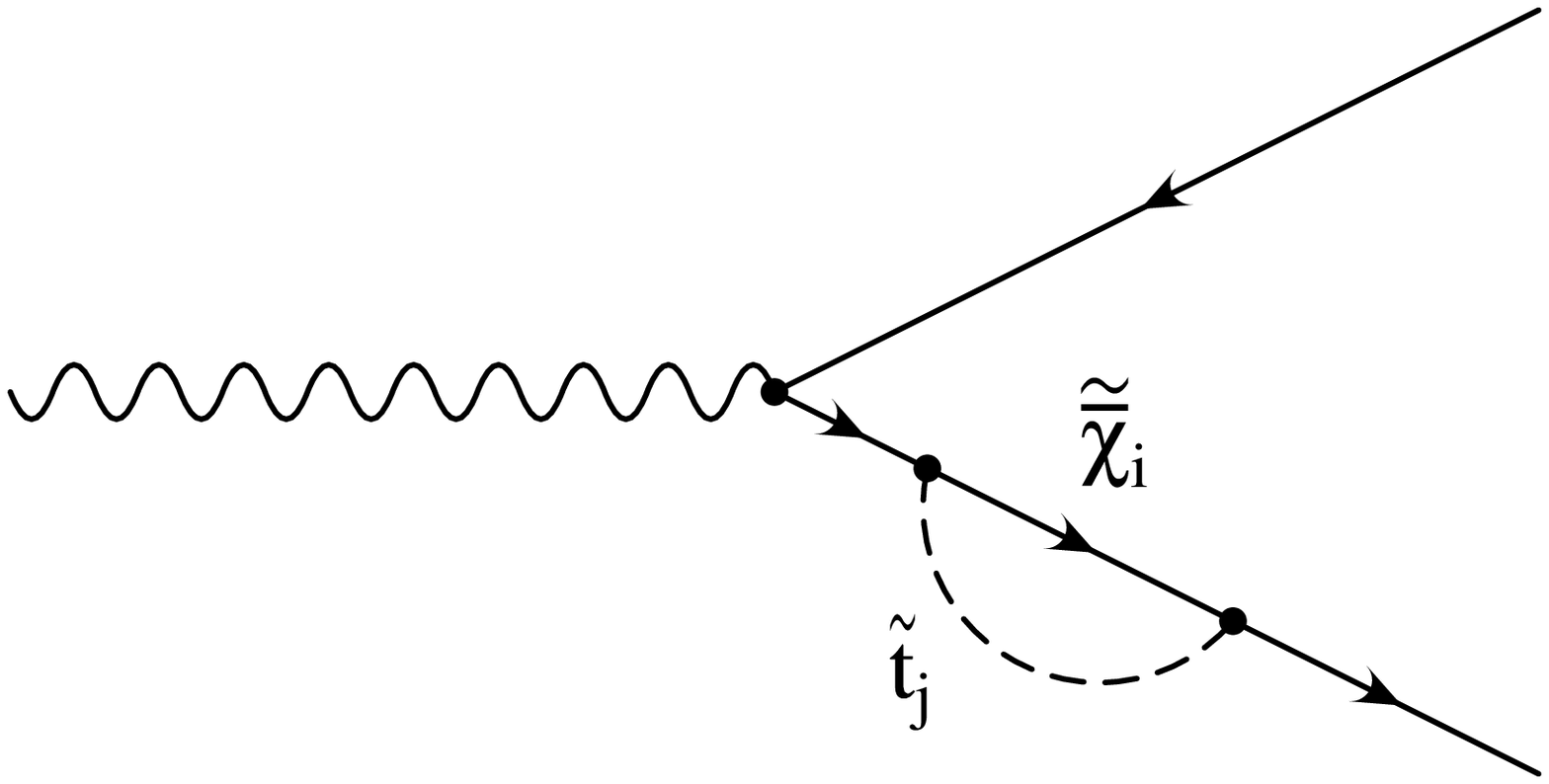,width=2.5in}
\psfig{figure=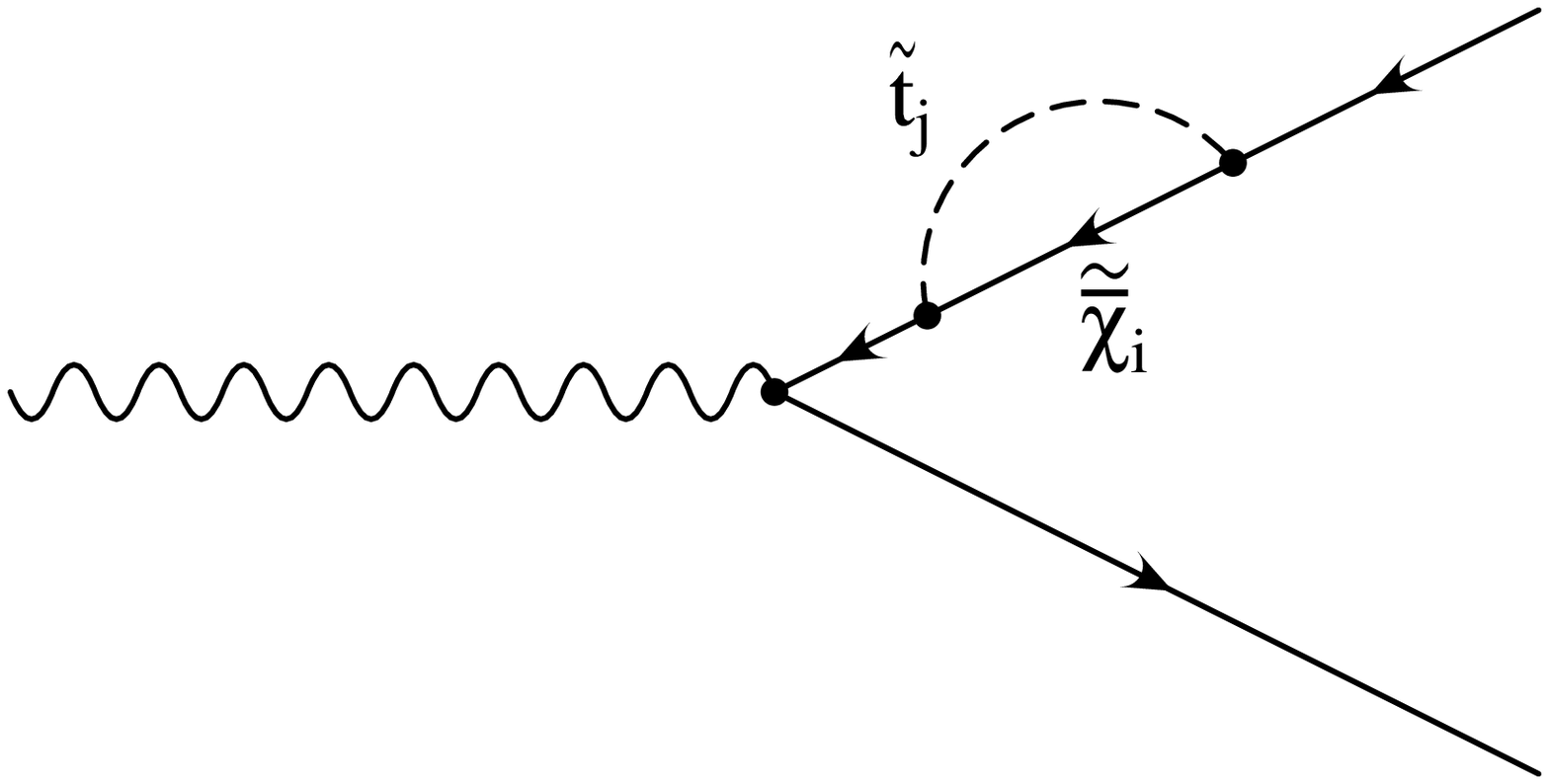,width=2.5in}}}
\begin{center}
\footnotesize{{\bf Figure 2a :} Self energy chargino and squark 
 corrections to the
$Zb\bar{b}$ vertex.}
\end{center}
%%%%%%%%%%%%%%%%%%%%%

%%%%%%%%%%%%%%%
%\vspace*{1in}
\centerline{\psfig{figure=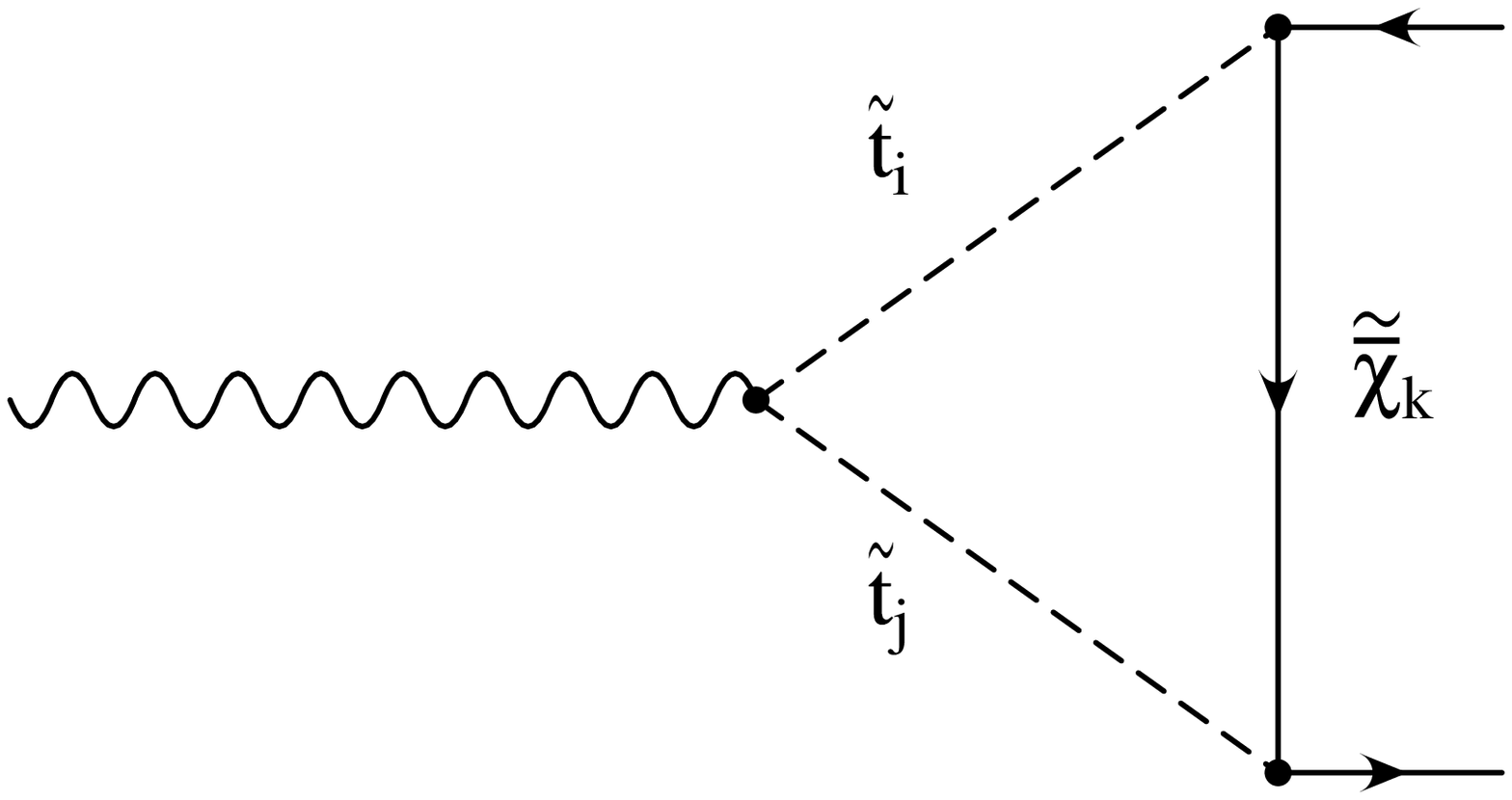,width=2.5in}}
%\begin{center}
%%%%%%%%%%%%%%%%%%%%
\vspace*{0.4in}
\centerline{\psfig{figure=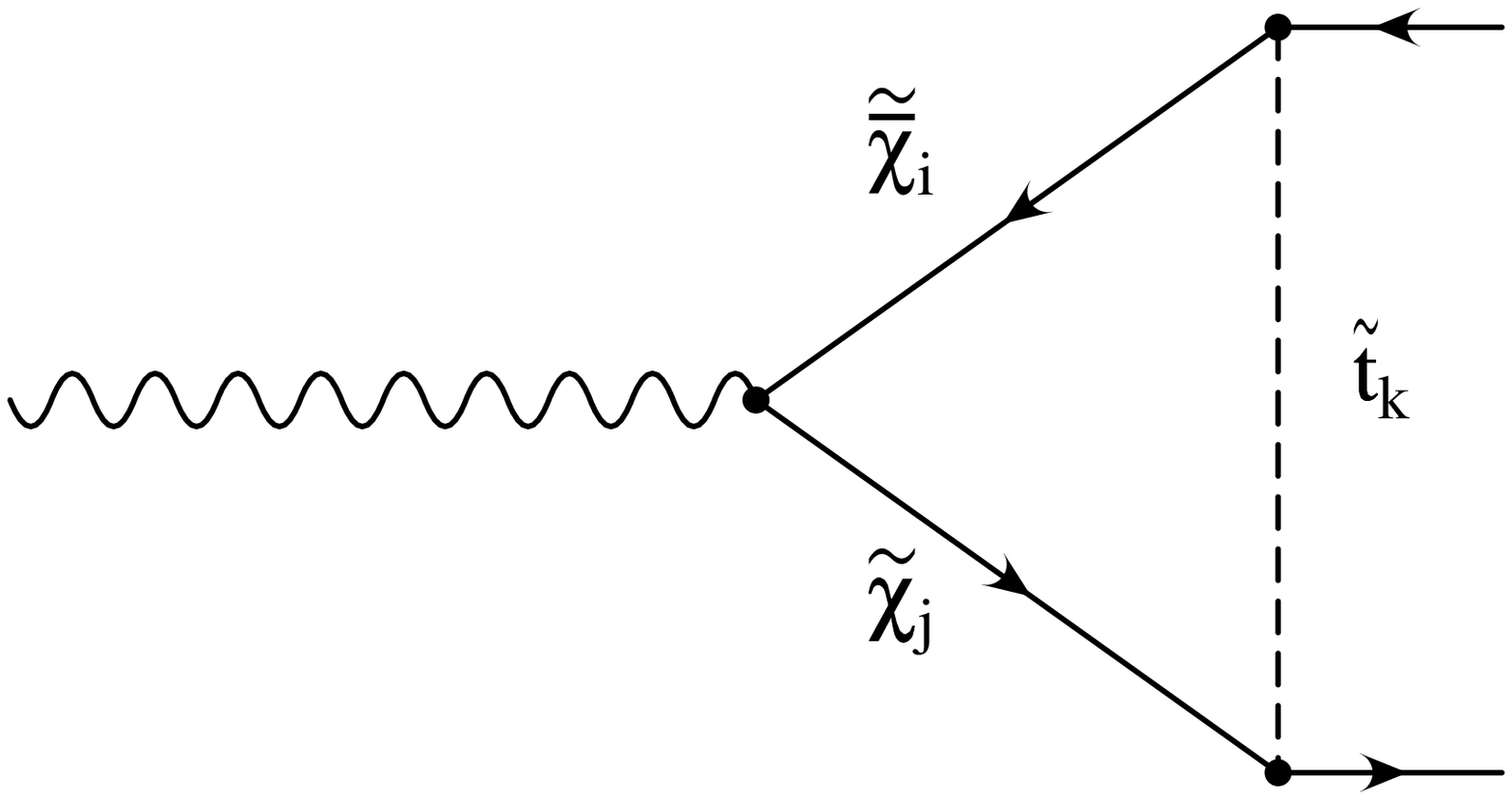,width=2.5in}}
\begin{center}
\footnotesize{{\bf Figure 2b :} Chargino and squark contributions to
the $Zb\bar{b}$ vertex.}
\end{center}
%%%%%%%%%%%%%%%%%

%%%%%%%%%%%%%%%%%%%%
%\vspace*{1in}
\centerline{\psfig{figure=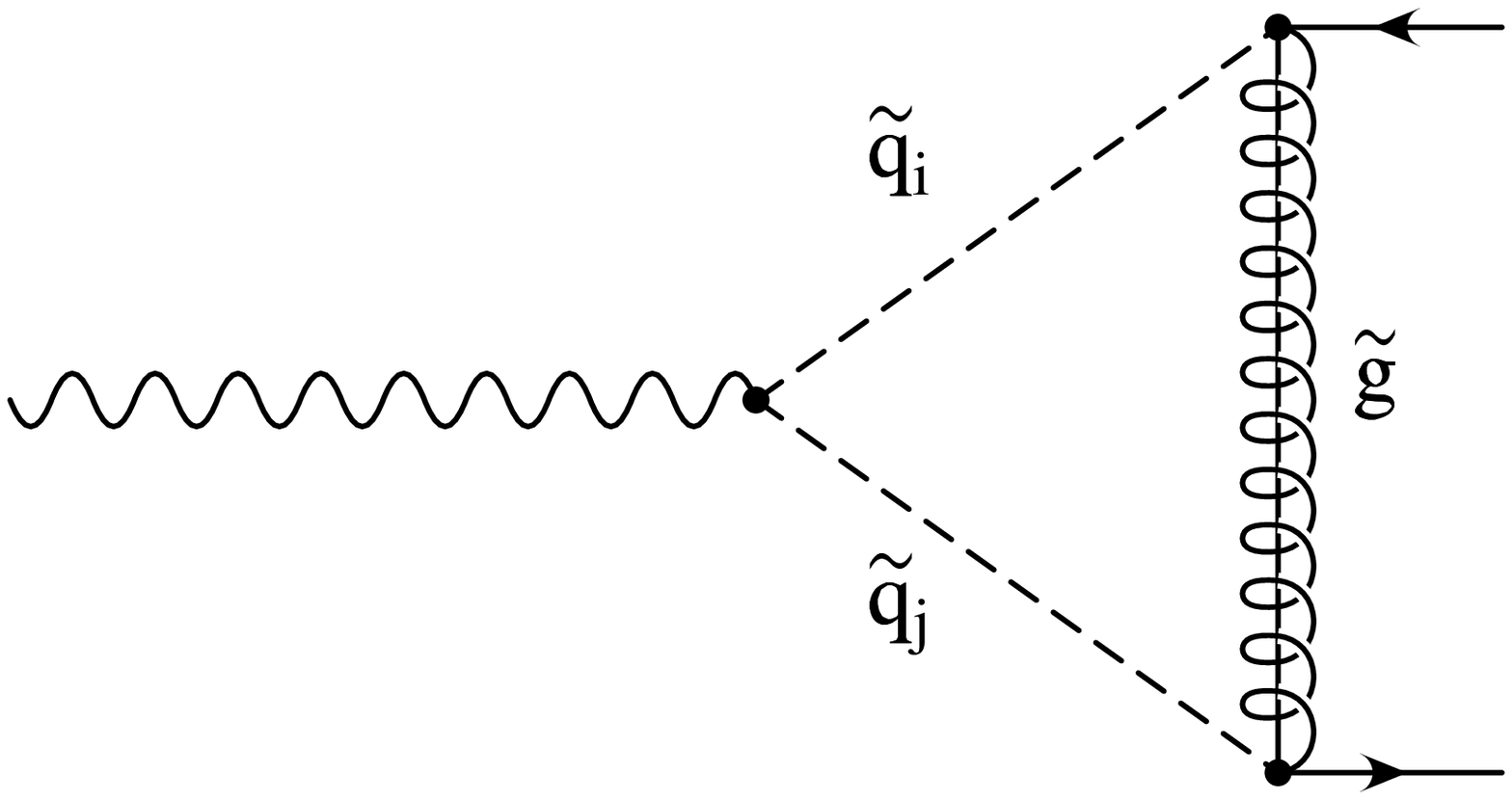,width=2.5in}}
\begin{center}
\footnotesize{{\bf Figure 2c :} Supersymmetric QCD corrections to
the $Zq\bar{q}$ vertex from gluino and squark contributions.}
\end{center}
%%%%%%%%%%%%%%%%%

%%%%%%%%%%%%%%%%%
\centerline{\hbox{\psfig{figure=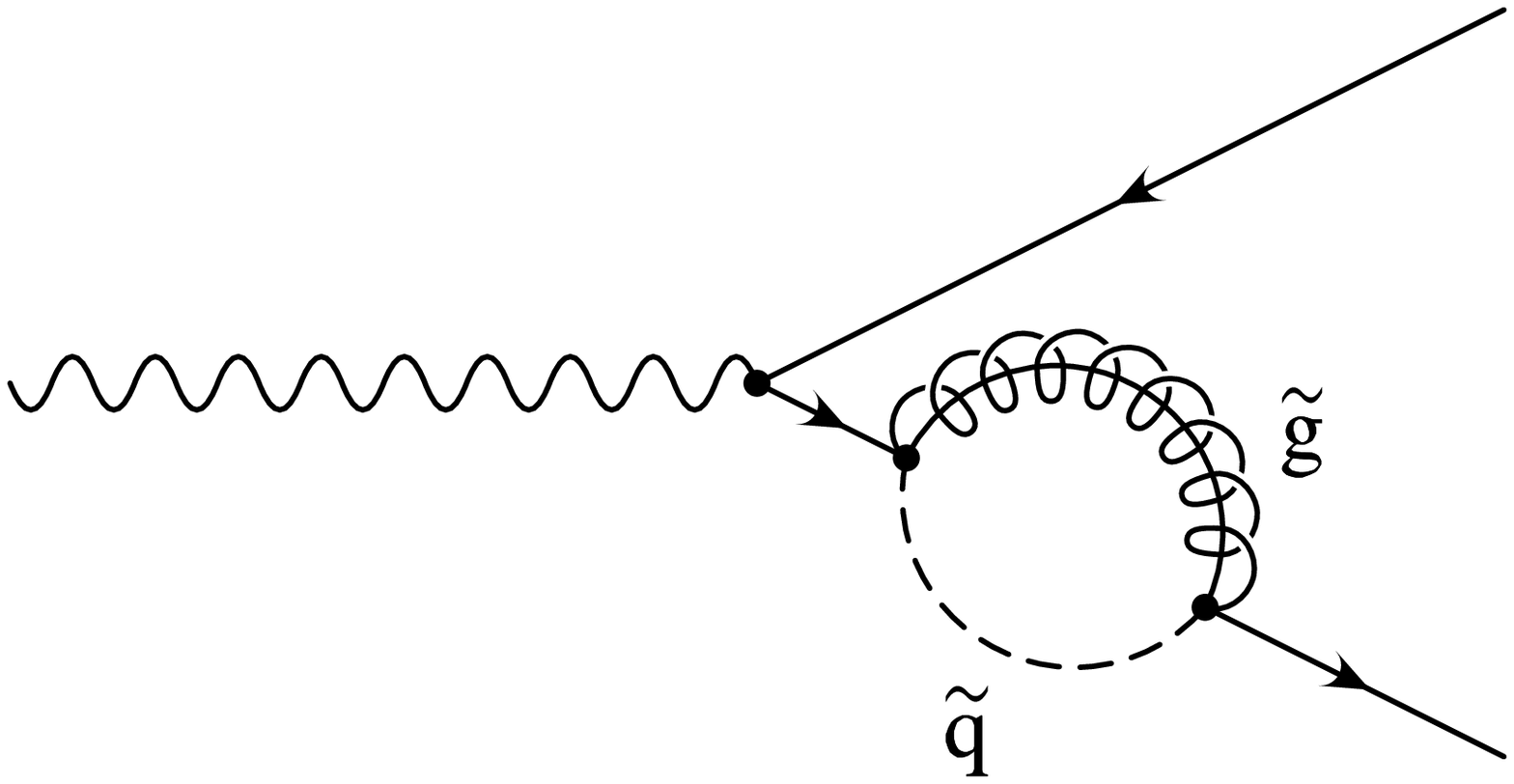,width=2.5in}
\psfig{figure=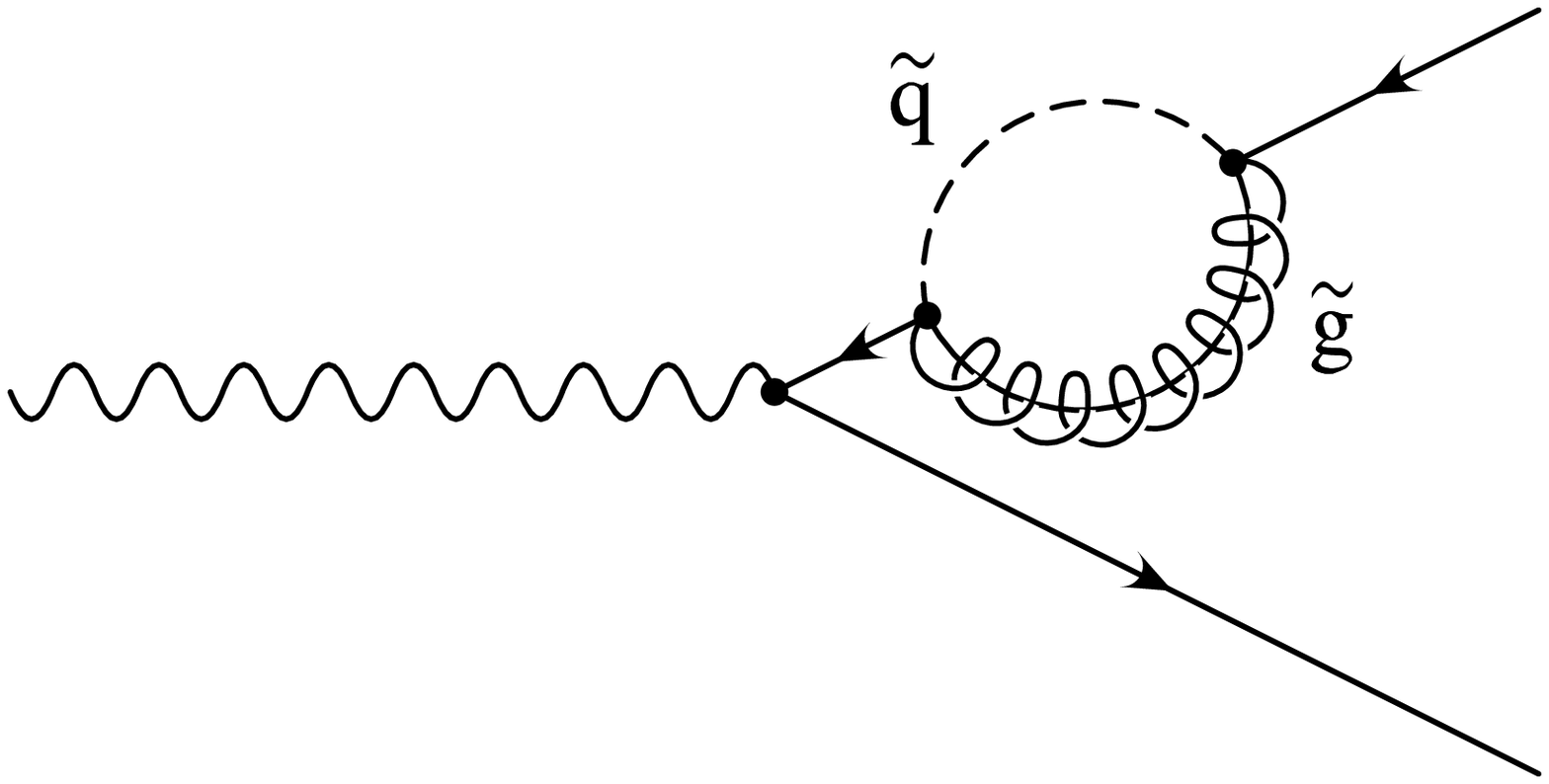,width=2.5in}}}
\begin{center}
\footnotesize{{\bf Figure 2d :} Self energy
gluino and squark contributions to the
$Zq\bar{q}$ vertex.}
\end{center}
%%%%%%%%%%%%%%%%%%%%%

\vspace*{0.25in}
%%%%%%%%%%%%%%%%%%%%%%%%%%%
\centerline{\input{sinem12+.ps}}
\begin{center}
\footnotesize{{\bf Figure 3 :}
 The effective weak mixing angle $s_l^2(M_Z)$ in comparison to the
weak mixing angle $\hat{s}^2(M_Z)$ versus $M_{1/2}$ when the
soft parameters $M_0$, $A_0$ vary in the indicated regions.
The width of each branch is due mainly to the variation on $M_0$,
for low $M_{1/2}<200$ GeV, and to the variation of the top mass
for $M_{1/2}>200$ GeV.
The effect of the variation of $A_0=0-900$ GeV and of $\tan\beta=5-28$ 
on $s_l^2$ is negligible.
The error bar show the measured value of $s_l^2=0.23152 \pm 0.00023$,
 obtained at LEP and SLD. 
 The MSSM value is in agreement with the LEP+SLD data for  the bulk of the values
in the soft parameter space.}
\end{center}
%%%%%%%%%%%%%%%%%%%%%%%%%%%

\vspace*{0.25in}
%%%%%%%%%%%%%%%%%%%%
\centerline{\input{sinall+.ps}}
\begin{center}
\footnotesize{{\bf Figure 4 :} The effective weak mixing angles 
$s^2_c$ and $s^2_b$ . In the region
 $M_{1/2} \rightarrow 900 \, GeV$, the two angles are
separated from each other. 
The dispersion of points around the central value for $M_{1/2}>200$ GeV
is due to the variation of the top mass. For the limiting behaviour 
to be more clearly exhibited, in this  figure
and in figures 3,7 and 8 we do not display the dispersion of points for
$M_{1/2}>800$ GeV.}
% and they agree with the LEP average value
%s^2_f=0.23199 \pm 0.00028$.}
\end{center}
%%%%%%%%%%%%%%%%%%%%%

\newpage

\vspace*{2.25in}
%%%%%%%%%%%%%%%%%%%%
\centerline{\input{mt+.ps}}
\begin{center}
\footnotesize{{\bf Figure 5 :} MSSM predictions for $s^2_l$ as a function
of $m_t$ for two different characteristic values of the
soft breaking parameters. The corresponding values of the light Higgs mass
and their errors due to the variation of $m_t$ are indicated.}
\end{center}
%%%%%%%%%%%%%%%%%%%%%

\newpage

\vspace*{2.25in}
%%%%%%%%%%%%%%%%%%%%
\centerline{\input{mwmt+.ps}}
\begin{center}
\footnotesize{{\bf Figure 6 :} MSSM predictions for physical mass of the
W-boson as a funcion of $m_t$ for the same inputs as in Figure 5.}
% Values around $M_0=M_{1/2}=300 \, GeV$ are
%preferred. }
\end{center}
%%%%%%%%%%%%%%%%%%%%%

\vspace*{0.25in}
%%%%%%%%%%%%%%%%%%%%
\centerline{\input{mw+.ps}}
\begin{center}
\footnotesize{{\bf Figure 7 :}  MSSM prediction for the mass
 of the W-gauge boson 
 as a function of the independent soft parameters $M_{1/2}$, $M_0$ and $A_0$.  
 The  experimental value $M_W=80.427 \pm 0.075$ ($M_W=80.405 \pm 0.089$) GeV
obtained at LEP (CDF,UA2,D{\O})
 is shown for comparison. }
\end{center}
%%%%%%%%%%%%%%%%%%%%%

\vspace*{0.25in}
%%%%%%%%%%%%%%%%%%%%
\centerline{\input{ae+.ps}}
\begin{center}
\footnotesize{{\bf Figure 8 :} The  left-right asymmetry $A_e$ in the
MSSM as a function
of $M_{1/2}$ when we vary  $M_0$, $A_0$, $\tan\beta$ and $m_t$.}
\end{center}
%%%%%%%%%%%%%%%%%%%%%

\vspace*{2in}
%%%%%%%%%%%%%%%%%%%%
\centerline{\input{all+.ps}}
\begin{center}
\footnotesize{{\bf Figure 9 :} Acceptable values in the $M_{1/2}$-$M_0$
plane according to the LEP+SLD data.
 The values of $\tan\beta$ and $A_0$ are taken 
 in the region $2-30$ and $0-900$ GeV, respectively. Only large
values of $M_{1/2}$ are acceptable.}
\end{center}
%%%%%%%%%%%%%%%%%%%%%


\begin{thebibliography}{99}

\bibitem{Lang}  see for instance, P. Langacker in {\it ``Precision Experiments,
Grand Unification and Compositeness"}, invited talk SUSY 95, Palaiseau
 France, May 1995, NSF-ITP-95-140, UPR-0683T.

\bibitem{LEP}
R. M. Barnett {\it al.}, \pr{54}{1}{1996}, 
and 1997 off-year partial update for the 1998 edition available on 
    the PDG WWW pages ({\tt URL: http://pdg.lbl.gov/}). 

%LEP Electroweak Working Group and the SLD Heavy Flavour Group
%CERN-PRE/96-183,  contribution
%to the 28th International Conference on High Energy
%Physics, Warsaw, Poland (1996).

\bibitem{altarelli}
G. Altarelli, R. Barbieri and F. Caravaglios, e-print {\tt hep-ph/9712368}.

\bibitem{sirlin}
G. Degrassi and A. Sirlin, \np{352}{342}{1991}.


\bibitem{gambino}
P. Gambino and A. Sirlin, \pr{49}{1160}{1994}.

%%%%%%%%%%%%%%%%%%%%%%%%%%%%%
\bibitem{all1}
W. J. Marciano and J. L. Rosner, \prl{65}{2963}{1990} and Erratum {\it ibid.}
{\bf 68}, 898 (1992); 

G. Degrassi, S. Fanchiotti and A. Sirlin, \np{351}{49}{1991}; 

B. A. Kniehl and A. Sirlin, \pr{47}{883}{1993}; 

S. Fanchiotti, B. A. Kniehl and A. Sirlin, \pr{48}{307}{1993}; 

B. A. Kniehl, presented at Tennessee International
Symposium on Radiative Corrections: 
Status and Outlook, Gatlinburg, TN, 27 Jun - 1 Jul 1994; 

A. Sirlin, \pl{348}{201}{1995}, addendum-{\it ibid} 
{\bf B352}, 498 (1995);

B. A. Kniehl and A. Sirlin, \np{458}{35}{1996}; 

P. Gambino, {\it Acta. Phys. Polon.} {\bf B27}, 3671 (1996);

G. Degrassi, P. Gambino and A. Sirlin, \pl{394}{188}{1997}.
%%%%%%%%%%%%%%%%%%%%%%%%%%%%%%%%%%%%

\bibitem{all2}
R. Barbieri, M. Beccaria, P. Ciafaloni, G. Curci and A. Vicere,
\pl{288}{95}{1992} ; \np{409}{105}{1993};

J. Fleischer, O. V. Tarasov and F. Jegerlehner, \pl{319}{249}{1993};

G. Buchalla and A. J. Buras, \np{398}{285}{1993};

L. Andeev et al, \pl{336}{560}{1994}, E: {\it {ibid}}.
{\bf{B349}} 597 (1995); 

K. G. Cheturkin, J. H. K{\"{u}}hn and M. Steinhauser,
\prl{75}{3394}{1995};

K. G. Cheturkin, J. H. K{\"{u}}hn and A. Kwiatkowski, in:
{\it {Precision Calculations for the Z resonance}}, CERN 95-03,eds.
D. Bardin, W. Hollik, G. Passarino, S. Peris, A. Santamaria,
CERN-TH-95-21 (1995);

G. Degrassi, P. Gambino and A. Vicini, \pl{383}{219}{1996}. 

%%%%%%%%%%%%%%%%%%%%%%%%%%%%%%%%%
\bibitem{djoua}
A. Djouadi and C. Verzegnassi, \pl{195}{265}{1987};

A. Djouadi, Nuovo Cimento {\bf{100A}} (357) 1988;

B. A. Kniehl, \np{347}{86}{1990};

A. Djouadi and P. Gambino, \pr{49}{3499 and 4705}{1994}, ibid.
{\bf{D51}} 218 (1995), E: {\it {ibid}}. {\bf{D53}} 4111 (1996). 

%%%%%%%%%%%%%%%%%%%%%%%%%%%
\bibitem{NHK}H. P. Nilles, {\it Phys. Rep. }
{\bf 110} (1984)1;\\
H. E. Haber and G. L. Kane,
{\it Phys. Rep. }{\bf 117} (1985)75;\\
 A. B. Lahanas and D. V. Nanopoulos, {\it Phys. Rep.}
{\bf 145} (1987)1.

%%%%%%%%%%%%%%%%%%%%%%%%%%%%%%%
\bibitem{polon}
M. Drees, K. Hagiwara, and A. Yamada, \pr{45}{1725}{1992}; 

G. Altarelli, R. Barbieri, and F. Caravaglios, \pl{314}{357}{1993}; 

P. Langacker and N. Polonski, \pr{47}{4028}{1993}; {\bf{D52}} 3081 (1995); 

N. Polonski, Report No. UPR-0641-T, e-print {\tt hep-ph/9411378};

D. Garcia and J. Sola, \pl{354}{335}{1995};

G.L. Kane, R.G. Stuart and J.D. Wells, \pl{354}{350}{1995};

P. H. Chankowski, Z. Plucienik and S. Pokorski, \np{439}{23}{1995}; 

P.H. Chankowski and S. Pokorski, \pl{366}{188}{1996};

P. H. Chankowski, Z. Plucienik, S. Pokorski and C. E. Vayonakis,
\pl{358}{264}{1995};

D. Garcia and J. Sol\`{a}, \mpl{9}{211}(1994);

D. Garcia, R.A. Jim\'{e}nez and J. Sol\`{a}, \pl{347}{309}{1995},
{\it ibid.} {\bf B347} 321 (1995), and Erratum {\it ibid}
 {\bf B351} 602 (1995);

D.M. Pierce and J. Erler, e-print {\tt hep-ph/9708374}, talk
presented at the 5th International Conference on Physics Beyond the
Standard Model, Balholm (1997);

P.H. Chankowski, J. Ellis and S. Pokorski, e-print {\tt hep-ph/9712234};

W.~de Boer {\it et al.}, e-print {\tt hep-ph/9712376};

D.M. Pierce and J. Erler, e-print {\tt hep-ph/9801238}.
%%%%%%%%%%%%%%%%%%%%%

%%%%%%%%%%%%%%%%%%%%%%%%%%%%%%%
\bibitem{Bagger}J. Bagger, K. Matchev, D. Pierce and R. Zhang, 
\np{491}{3}{1997}.

\bibitem{Rociek}P. Chankowski et al, \np{417}{101}{1994};

P. Chankowski, S. Pokorski and J. Rosiek, \np{423}{437}{1994}.

\bibitem{Hollik1}
W. de Boer, A. Dabelstein, W. Hollik, W. Mosle, U. Schwickerath,
\zp{75}{627}{1997}; \\
W. Hollik in: {\it { Electroweak Precision Observables in the MSSM
and Global Analysis of Precision Data}}, Talk at the International
workshop on Quantum Effects in the MSSM, Barcelona 9-23 September 1997.

\bibitem{finnell}
M. Boulware and D. Finnell, \pr{44}{2054}{1991}.

\bibitem{kolda}
J.D. Wells, C. Kolda and G.L. Kane, \pl{338}{219}{1994};

M. Drees, R.M. Godbole, M. Guchait, S. Raychaudhuri, D.P. Roy,
\pr{54}{5598}{1996};

P.H. Chankowski and S. Pokorski, \np{475}{3}{1996}.


\bibitem{Tamvakis}J. Ellis, J. Hagelin, D.V. Nanopoulos and K. Tamvakis,
\pl{125}{1983}{275};

L. E. Ibanez and G. G. Ross, \pl{110}{215}{1982};

L. Alvarez-Gaume, J. Polchinski and M. Wise, \np{221}{495}{1983};

%J. Ellis, A.B. Lahanas, D.V. Nanopoulos and K. Tamvakis,
%\pl{134}{429}{1983}.

\bibitem{Siegel}
W. Siegel, \pl{84}{193}{1979}; 

 D. M. Capper, D. R. T. Jones and
P. van Nieuwenhuizen,   \np{167}{479}{1980}; 

I. Antoniadis, C. Kounnas and K. Tamvakis, \pl{119}{377}{1982};

S. P. Martin and M. T. Vaughn, Phys. Lett. B318(1993)331.


\bibitem{Martin}
I. Jack, D. R. T. Jones, S. P. Martin, M. T. Vaughn and
Y. Yamada, \pr{50}{5481}{1994}.




%\bibitem{NHK}H. P. Nilles, {\it Phys. Rep. }
%{\bf 110} (1984)1;\\
%H. E. Haber and G. L. Kane,
%{\it Phys. Rep. }{\bf 117} (1985)75;\\
% A. B. Lahanas and D. V. Nanopoulos, {\it Phys. Rep.}
%{\bf 145} (1987)1.


\bibitem{dedes2}
A. Dedes, A. B. Lahanas, J. Rizos and K. Tamvakis, \pr{55}{2955}{1997}.

\bibitem{Peskin}
M. E. Peskin and D. V. Schroeder, {\it An Introduction to Quantum
Field Theory}, Addison-Wesley, 1995.


%\bibitem{LEPII}
%The LEP Collaborations ALEPH, DELPHI, L3, OPAL, the LEP
%Electroweak Working Group and the SLD Heavy Flavour 
%Working Group, CERN-PRE/96-183,  contributed
%to the 28th International Conference on High Energy
%Physics, Warsaw, Poland (1996).


\bibitem{Pokorski}P.H. Chankowski, A. Dabelstein, W. Hollik, 
W.M. Mosle, S. Pokorski and J. Rosiek, \np{417}{101}{1994}.

\bibitem{passarino}
G. Passarino and M. Veltman, \np{160}{151}{1979}.

\bibitem{ahn}
C. Ahn, B. Lynn, M. Peskin and S. Selipsky, \np{309}{221}{1988}.

\bibitem{2loop}
S. P. Martin and M. T. Vaughn, \pr{50}{2282}{1994};

Y. Yamada, \pr{50}{3537}{1994};

I. Jack and D. R. T. Jones, \pl{333}{372}{1994}.

\bibitem{olden}
G. J. van Oldenborgh, {\it {Comput. Phys. Commun. }} {\bf {66}} 
1 (1991).

\bibitem{Tata}
X. Tata, Lectures given at 9th Jorge
Andre Swieca Summer School: Particles and Fields, Sao Paulo, Brazil, 
16-28 Feb 1997, e-print {\tt hep-ph/9706307}.

\bibitem{Dedes3}
A. Dedes and K. Tamvakis, \pr{56}{1496}{1997}.

\bibitem{top}

D{\O} Collaboration: S. Abachi {\it et al.}, \prl{79}{1197}{1997};

CDF Collaboration: F. Abe {\it et al.}, \prl{79}{1992}{1997}.

\bibitem{Hollik2}
W. Hollik in {\it {Renormalization of the Standard Model}}, published in
 {\bf {"Precision Tests of the Standard Model"}}, World Scientific, ed.
 P. Langacker.

\bibitem{sqcdrefs}
L. Clavelli, P.W. Coulter and L.R. Surguladze, e-print {\tt hep-ph/9712547}.

The two loop QCD corrections to the scalar quark contributions to the
$\rho$ parameter is the subject of:

A. Djouadi, P. Gambino, S. Heinemeyer, W. Hollik, C. J\"{u}nger and
G. Weiglein, \prl{78}{3626}{1997};

A. Djouadi, e-print {\tt hep-ph/9710440};

A. Djouadi, P. Gambino, S. Heinemeyer, W. Hollik, C. J\"{u}nger and
G. Weiglein, e-print {\tt hep-ph/9710438};

G. Weiglein, 
Talk given at 21st
International School of Theoretical Physics (USTRON 97), 
Ustron, Poland, 19-24 Sep 1997, e-print {\tt hep-ph/9711254}.

\bibitem{Pok3}
P. Chankowski and S. Pokorski , hep-ph/9707497, to appear 
in ``Perspectives in supersymmetry" edited by G.L. Kane,
World Scientific, 1997. 

\end{thebibliography}
\end{document}